\documentclass[aps,preprint,a4paper,showpacs,showkeys,superscriptaddress]{revtex4-1}

\usepackage{latexsym}
\usepackage{amsmath,amssymb}
\usepackage{graphicx}
\usepackage{subfigure}
\usepackage{xcolor}

\usepackage{hyperref}     
\hypersetup{colorlinks,%
  citecolor=blue,%
  linkcolor=cyan,%
  pdftex}

\begin{document}


\title{The effective Tolman temperature in curved spacetimes}

\author{Wontae Kim}%
\email[]{wtkim@sogang.ac.kr}%
\affiliation{Department of Physics, Sogang University, Seoul 04107, South Korea}%

\begin{abstract}
We review a recently proposed effective Tolman temperature and present its applications to
various gravitational systems.
In the Unruh state for the evaporating black holes,
the free-fall energy density is found to be negative divergent at the horizon,
which is in contrast to the conventional calculations performed in the Kruskal coordinates.
We resolve this conflict by invoking that the Krukcal coordinates could be no longer
proper coordinates at the horizon.
In the Hartle-Hawking-Israel state,
despite the negative finite proper energy density at the horizon,
the Tolman temperature is divergent there due to the infinite blueshift of the Hawking temperature.
However, a consistent Stefan-Boltzmann law with the Hawking radiation
shows that the effective Tolman temperature is eventually finite everywhere
and the equivalence principle is surprisingly restored at the horizon.
Then, we also show that the firewall necessarily emerges out of the Unruh vacuum,
so that the Tolman temperature in the evaporating black hole is naturally divergent
due to the infinitely blueshifted {\it negative ingoing flux} crossing the horizon,
whereas the outgoing Hawking radiation characterized by the effective Tolman temperature
indeed originates from the quantum atmosphere, not just at the horizon.
So, the firewall and the atmosphere for the Hawking radiation turn out to be compatible,
once we discard the fact that the Hawking radiation in the Unruh state originates from the infinitely blueshifted outgoing excitations at the horizon.
Finally, as a cosmological application, the initial radiation energy density
in warm inflation scenarios has been assumed to be
finite when inflation starts.
We successfully find the origin of the non-vanishing initial radiation energy density
in the warm inflation by using the effective Tolman temperature.

\end{abstract}

\maketitle


\section{Introduction}
\label{sec:intro1}
Hawking radiation is of relevance to
not only information loss problem in
the theory of quantum gravity~\cite{Hawking:1974rv,Hawking:1974sw,Hawking:1976ra} but
also black hole complementarity~\cite{Susskind:1993if,
  Susskind:1993mu, Stephens:1993an}. In particular, the latter implies
that there are no
contradictory physical observations between a freely falling observer
and a rest observer since the two descriptions are complementary.
The presence of Hawking radiation indicates that the rest observer at
infinity sees the flux of particles. However, it has been shown
that ``a geodesic detector near the horizon will not see the Hawking
flux of particles"~\cite{Unruh:1976db}, while the infalling
negative energy flux
can exist near the horizon \cite{Unruh:1977ga}.

Note that the energy densities in the Unruh state \cite{Unruh:1976db} and
the Hartle-Hawking-Israel state \cite{Hartle:1976tp,Israel:1976ur}
are finite on the future horizon, while
it is divergent in the Boulware state  \cite{Boulware:1974dm}.
Explicitly, for the black hole in the Unruh state,
the energy-momentum tensors were calculated
at the bifurcation two-sphere
in virtue of the vanishing affine connections \cite{Candelas:1980zt}.
This calculation was in turn extended to the future horizon
by taking into account of a symmetry argument for the infinite time,
 and the finite energy density was eventually obtained
on the future horizon when the observer is dropped from rest
at the future event horizon without any journey.
However, it was claimed that
the freely falling observer finds the firewall at the
event horizon and burns up because of high energy
quanta beyond the Planckian scale~\cite{Almheiri:2012rt}
after the Page time \cite{Page:1993wv}
in the semiclassical approximations.
Conventionally, the equivalence
principle tells us that a freely falling observer can not see any
radiation. This fact is based on the classical argument of
locality but it may not be true in quantum regime
such that the freely falling observer can find non-trivial quantum-mechanical
radiation and temperature~\cite{Brynjolfsson:2008uc, Greenwood:2008zg,
Barbado:2011dx, Smerlak:2013cha, Smerlak:2013sga,Kim:2013caa}.
Anyway, the presence of the firewall has something to do with the failure
of the equivalence principle or
breakdown of semiclassical physics at macroscopic distance from the horizon,
which eventually makes black hole complementarity incomplete.
Subsequently, much attention has
been paid to the firewall issue and many authors argued it pro and con \cite{Hossenfelder:2012mr,Bousso:2012as, Nomura:2012sw,
  Page:2012zc, Kim:2013fv, Giddings:2013kcj,
  Almheiri:2013hfa,Bousso:2012as,Nomura:2012sw,Page:2012zc,Giddings:2012gc,
JACOBSON:2013ewa,Kim:2013fv,Maldacena:2013xja, Hotta:2013clt,Page:2013mqa,Freivogel:2014dca,Saini:2015dea,Hutchinson:2013kka,
Braunstein:2009my}.

Interestingly, in the Unruh state for the evaporating black holes,
the free-fall energy density is found to be negatively divergent at the horizon \cite{Visser:1996ix,Eune:2014eka,Gim:2014swa},
which might be a signal of the firewall. But it is in contrast to the above
conventional calculations performed in the Kruskal coordinates \cite{Candelas:1980zt}.
So, we would like to elaborate the issue clearly in Sec. \ref{sec:ff.frame}, and
resolve the conflict by invoking that the Krukcal coordinates could be no longer
proper coordinates at the horizon in Sec. \ref{sec:intro3}.

Now, in the Hartle-Hawking-Israel state,
the free-fall energy density and pressure are finite
at the horizon
even though the Tolman temperature is infinite at the horizon \cite{Page:1982fm}.
It implies that
the Stefan-Boltzmann law
relating the proper energy density to the local temperature might be nontrivial.
Moreover, the energy density at the horizon is negative in this vacuum,
which indicates that the negative energy density should be related to the positive
temperature non-trivially. At first sight, it seems to be impossible to resolve this problem
if the conventional
Stefan-Boltzmann law persists.
We will derive the effective Tolman temperature consistently
in order to resolve the present issue in the regime of the semiclassical quantum field theory and thermodynamics
in Sec. \ref{sec:intro4}.
To shed light on the essential feature of our formulation with exact solvability,
we adopt the two-dimensional approach to the problem.
First of all, we note that the energy-momentum tensor of matter fields on the classical background metric receives
semiclassical quantum corrections which give rise to the trace anomaly \cite{Deser:1976yx}.
Note that the conventional Tolman temperature is correct only for
the traceless case \cite{Tolman:1930zza,Tolman:1930ona}, so that it should
be generalized semiclassically for a consistent formulation
when Hawking radiation is involved, since Hawking radiation is
indeed related to the trace anomaly of matter fields \cite{Christensen:1977jc}.
To get the consistent local proper temperature of the black hole,
the traceless condition of the energy-momentum tensor should be released {\it ab initio}
\cite{Gim:2015era}.
Next, the above issue will be extended to the case of
the four-dimensional Schwarzschild black hole \cite{Eune:2015xvx}, where
the renormalized stress tensor is no more isotropic in Sec. \ref{sec:intro5}.

Returning back to the issue on the evaporating black hole,
Unruh attained a startling conclusion that the Hawking radiation
appears in the absence of the outgoing flux at the horizon
\cite{Unruh:1976db} and also
showed that the process of thermal particle creation is low energy behavior
and the highest frequency mode does not matter for the thermal emission by
using a modification of the dispersion relation in a sonic black hole numerically \cite{Unruh:1994zw}.
Moreover, it was shown that the
effective blueshift of the outgoing Hawking radiation
remains finite
by averaging out the Tolman factor outside the horizon in terms of the Poisson distribution
\cite{Casadio:2002dj}.
All these imply that the Hawking radiation originates from a macroscopic distance
outside the horizon.
Interestingly, Israel claimed that the Hawking radiation can be retrieved by an alternative scenario
that a positive outward flux at the horizon is interpreted as the negative influx
without recourse to a pair creation scenario \cite{Israel:2015ava}.
Recently, there was a refined question concerning the origin of the Hawking radiation in the Unruh vacuum
by Giddings \cite{Giddings:2015uzr},
where the evidence concerns the three relevant points that are: (1) the effective emitting area of Hawking radiation is
considerably larger than the size defined by the area of the black hole~\cite{Page:1976df},
(2) there is no outgoing flux at the horizon in the Unruh vacuum
and there should appear a transition from the ingoing to the outgoing flux
over a large quantum region, and
(3) the size of wave length of a thermal Hawking particle is larger than
the horizon size.
One of the essential ingredients is that
the transition from the ingoing to the outgoing flux should appear over the atmosphere of
the quantum region outside the horizon
without resort to the firewall.
Moreover, the importance of
the atmosphere
was also emphasized in connection with the nonviolent scenarios
for the information loss paradox~\cite{Giddings:2006sj}.
However, the existence of the firewall seems obvious
from the fact that the Tolman temperature defined in the Unruh vacuum
is infinite on the horizon \cite{Tolman:1930zza},
which is indeed due to the infinite blueshift of the Hawking temperature there.
In other words, the thermal Hawking particles at infinity
might be ascribed to the infinitely blueshifted outgoing radiation at the horizon or very
near the horizon.
This argument, as mentioned above, would not be reliable, since the Unruh vacuum
does not admit any outgoing flux on the horizon semiclassically \cite{Unruh:1976db}.
So it is not likely to get the firewall from the outgoing flux, and thus
it might be tempting to conclude
that the firewall is incompatible with the Unruh vacuum.
In these regards, the origin of Hawking radiation and the reason for the
existence of the firewall as well as their relationship
still seem to be equivocal in spite of many efforts.
In Sec. \ref{sec:intro6},
we will show that the firewall necessarily emerges out of the Unruh vacuum,
so that the Tolman temperature in the evaporating black hole should be divergent
due to the infinitely blueshifted negative ingoing flux crossing the horizon rather than the outgoing flux,
whereas the outgoing Hawking radiation characterized by the effective Tolman temperature
indeed originates from the quantum atmosphere, not just at the horizon.
So, the firewall and the atmosphere for the Hawking radiation turn out to be compatible,
once we discard the fact that the Hawking radiation in the Unruh state originates from the infinitely blueshifted outgoing excitations at the horizon \cite{Kim:2016iyf}.

On the other hand, in the big bang cosmology,
inflation is an elegant solution to the intriguing problems such as the horizon and flatness problems \cite{Starobinsky:1980te, Sato:1980yn, Guth:1980zm}.
It also generates the perturbations
which are the origin of the spectrum of primordial gravitational waves \cite{Starobinsky:1979ty}, the cosmic microwave background (CMB) radiation
and the large scale structure of our universe \cite{Mukhanov:1981xt, Guth:1982ec, Hawking:1982cz, Starobinsky:1982ee, Bardeen:1983qw}.
The standard inflation, in particular, a chaotic inflation \cite{Linde:1981mu, Linde:1983gd}
is driven by scalar fields of the so-called inflaton.
 The inflationary expansion lays the universe in a supercooled phase,
and thereafter the universe is heated by assuming the reheating process.
In order to attain the explicit reheating process responsible for the graceful exit problem,
a wide variety of mechanisms of interest
have been studied
\cite{Abbott:1982hn, Albrecht:1982mp, Kofman:1994rk, Kofman:1997yn, Greene:1997ge, Allahverdi:2010xz, Amin:2014eta}.
In contrast to the assumption of the supercooled universe after inflation,
there has been another elegant way to approach this issue, that is, a warm inflation scenario without
reheating process \cite{Berera:1995ie, Berera:1996nv}.
The interactions of the inflaton and radiation are inevitable during inflation
via a damping term describing the decay rate of the inflaton into other fields,
and thus no large scale reheating is necessary,
where the curvature perturbations are generated
by a larger thermal fluctuation rather than a quantum fluctuation \cite{Berera:1995wh, Taylor:2000ze, Hall:2003zp}. In the framework of the warm inflation scenario,
we will get the initial non-vanishing radiation energy density nicely
by using the effective Tolman temperature from
the thermodynamic point of view in Sec. \ref{sec:intro7}.
If the initial radiation energy density were not zero,
then the radiation energy density at the initial stage of inflation
should be thermodynamically originated before inflation.
For our purpose, we will assume that the radiation and inflaton are in thermal equilibrium
in order to use thermodynamic relations consistently, and
more importantly treat the inflaton as an equal footing with the radiation thermodynamically.
Consequently, we shall find that the usual Stefan-Boltzmann law which is only valid in
cases of the traceless energy-momentum tensor should be modified effectively
because of the temperature-dependent effective potential
related to the non-vanishing trace of the
energy-momentum tensor \cite{Gim:2016uvv}.
The effective Tolman temperature tells us that
the radiation energy density in the warm inflation scenario starts from zero with the GUT temperature as an initial condition of our universe, and then it increases and becomes finite, which eventually gives the adequate initial radiation energy density
for warm inflation.
Finally, the summary will be given in Sec. \ref{sec:discussion8}.

\section{Proper energy density at the event horizon}
\label{sec:ff.frame}
In this section,
we are going to study the quantum-mechanical energy densities
 measured by the
freely falling observer on the two-dimensional Schwarzschild
black hole background. The trace anomaly for massless scalar fields
will be employed to calculate the energy-momentum tensors along with
covariant conservation law of the energy-momentum tensors. Then,
the energy density will be characterized by three states;
the Boulware \cite{Boulware:1974dm},
Unruh \cite{Unruh:1976db}, and Hartle-Hawking-Israel states \cite{Israel:1976ur,Hartle:1976tp} in order to investigate what state is
relevant to the infinite energy density at the horizon.  If there
exists such a non-trivial energy density at the horizon, then this fact will be tantamount to
the failure of no drama condition which has been one of the
assumptions for black hole complementarity.

\subsection{Free-fall frame}
Let us start with the two-dimensional Schwarzschild black hole
governed by~\cite{Unruh:1976db,Christensen:1977jc},
\begin{equation} 
  ds^2 = -f(r)dt^2 +\frac{1}{f(r)} dr^2, \label{metric}
\end{equation}
where the metric function is given by $f(r)=1-2M/r$ and the horizon is
defined at $r_{\rm H}= 2M$. Solving the geodesic equation for the
metric~\eqref{metric}, the proper velocity of a particle can obtained
as~\cite{Ford:1993bw}
\begin{align}
  u^\mu = \left(\frac{dt}{d\tau}, \frac{dr}{d\tau}\right) =
  \left(\frac{k}{f(r)}, \pm \sqrt{k^2 - f(r)} \right), \label{u:k}
\end{align}
where \(\tau\) and \(k\) are the proper time and the constant of
integration, respectively.  The $k$ can be identified
with the energy of a particle per unit mass for \(k > 1\),
which can be written as \(k = 1/\sqrt{1-v^2}\) with
\(v=dr/dt\).
In this case, the motion of the particle is unbounded,
so that the particle lies in the range of \(r\ge r_{\rm H}\).
For \(0 \le k \le 1\), the motion of the particle is bounded such that there is a
maximum point \(r_{\rm max}\) where  the particle lies in the range
of \(~r_{\rm H} \le r \le r_{\rm max}\).
We will consider a freely falling frame starting at
$r_s= r_{\rm max}$ with the zero velocity toward the black hole, which
can be shown to be the latter case by identifying $k =\sqrt{f(r_s)}$ in Eq.~\eqref{u:k},
and thus the proper velocity of a
freely falling observer can be written as
\begin{equation} 
  \label{eq:u}
  u^\mu = \left(\frac{dt}{d\tau}, \frac{dr}{d\tau}\right)
  = \left(\frac{\sqrt{f(r_s)}}{f(r)}, -\sqrt{f(r_s)-f(r)}\right).
\end{equation}
If the observer starts to fall into the black hole at the
spatial infinity, then $f(r_s)=1$ while
$f(r_s)=0$ for the observer to fall into the black hole just at the horizon.
Then, the radial velocity with respect to the Schwarzschild time
becomes $v = -f(r)\sqrt{f(r_s) - f(r)} / \sqrt{f(r_s)}$ which
vanishes both at the initial free-fall position and
the horizon, and the maximum
speed occurs at $r= 6M r_s/(4M+r_s)$.  The
proper time from $r_s$ to $r_{\rm H}$ is also obtained as
\begin{equation}
  \label{propertime}
  \tau = 2M \frac{\sqrt{f(r_s)(1-f(r_s))} + \sin^{-1} \sqrt{f(r_s)}}{(1-f(r_s))^{3/2}},
\end{equation}
which is finite except for the case of the free-fall at the asymptotic
infinity.
So, it would take a finite proper time to reach  the
event horizon when the free-fall begins at a finite distance.

In the light-cone coordinates defined by $\sigma^{\pm}=t \pm r^*$ through
$r^*=r+2M \ln(r/M-2)$ the proper velocity
\eqref{eq:u} can be written as
\begin{align}
  u^+ &=\frac{1}{\sqrt{f(r_s)}+\sqrt{f(r_s)-f(r)}}, \label{u:+} \\
  u^- &=\frac{\sqrt{f(r_s)} + \sqrt{f(r_s)-f(r)}}{f(r)}, \label{u:-}
\end{align}
where  $u^\pm=u^t
\pm u^r/f(r)$,
and the energy-momentum tensors are expressed as~\cite{Christensen:1977jc}
\begin{align}
  \langle T_{\pm \pm} \rangle&= -\frac{N}{48 \pi} \left(\frac{2M
      f(r)}{r^3} + \frac{M^2}{r^4} \right) + \frac{N}{48}t_{\pm},
  \label{T:++}   \\
  \langle T_{+-} \rangle &= -\frac{N}{48\pi}
  \frac{2M}{r^3}f(r), \label{eq:Tpm} 
\end{align}
where $N$ is the number of massless scalar fields and $t_{\pm}$ are
functions of integration to be determined by boundary conditions.
The two component covariant conservation law and the one sinlge trace equation
determine the explicit form of the three component energy-momentum tensor
with the two unknowns.

Now, the energy density measured by the freely falling observer can be
calculated as~\cite{Ford:1993bw, Poisson:1990eh},
\begin{equation}
  \label{energy}
  \epsilon = \langle T_{\mu \nu} \rangle u^\mu u^\nu
\end{equation}
by using the proper velocity and the energy-momentum tensor.
In connection with Hawking
radiation, the fields are quantized on the background metric
in such a way that non-trivial radiation will appear and the energy
density~\eqref{energy} will not vanish even in the freely falling frame.
Substituting Eqs.~\eqref{u:+}, \eqref{u:-}, \eqref{T:++} and
\eqref{eq:Tpm} into~\eqref{energy}, the energy density
can be
expressed as
\begin{align}  
  \label{eq:ed}
  \epsilon(r|r_s) &= -\frac{N}{48 \pi r^4 f(r)} \Bigg[ 8Mrf(r_s) +
  4M^2 \left(\frac{f(r_s)}{f(r)} - \frac{1}{2}\right)  - \pi r^4\left(\sqrt{\frac{f(r_s)}{f(r)}} -
    \sqrt{\frac{f(r_s)}{f(r)}-1}\right)^2t_+ \notag \\
&\quad -\pi r^4\left(\sqrt{\frac{f(r_s)}{f(r)}} +
    \sqrt{\frac{f(r_s)}{f(r)}-1}\right)^2t_-\Bigg],
\end{align}
which is reduced to
\begin{align}\label{eq:edr0}
  \epsilon\ (r_s|r_s)= & -\frac{N}{48 \pi r_s^4 f(r_s)} [
  8Mr_sf(r_s)+2M^2 -\pi r_s^4(t_++t_-) ],
\end{align}
at the special limit of $r=r_s$.
Next, let us investigate some characteristics
of the free-fall energy density measured in
the Boulware, Unruh, and Hartle-Hawking-Israel states, respectively.

 \subsection{Boulware state}
 \label{boulware}

\begin{figure}[hbt]
  \begin{center}
  \includegraphics[width=0.6\textwidth]{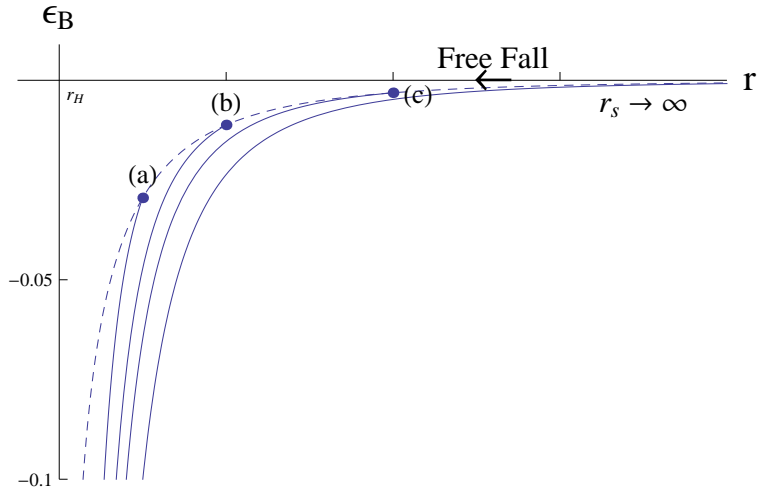}
  \end{center}
  \caption{The energy densities in the Boulware state are plotted by
    choosing as $N=12, M=1$. They are
 always negative and independent of the initial free-fall positions.
    The solid curves represent  $\epsilon_B \ (r|r_s)$,
    and the dotted curve is for $\epsilon_B\ (r_s|r_s)$.
    The three energy densities at $r_s$ are denoted
    by the black dots (a), (b), and (c).}
  \label{fig:EB}  
\end{figure}

The Boulware state is obtained by choosing $t_\pm=0$, where the
energy density~\eqref{eq:ed} reads as
\begin{equation}
  \epsilon_B(r|r_s) = -\frac{NM^2}{12 \pi r^4
    f(r)} \left[\frac{2 r f(r_s)}{M} + \frac{f(r_s)}{f(r)} -
    \frac{1}{2} \right], \label{energy:B}
\end{equation}
which is always negative.  So the
freely falling observer encounters more and more negative energy
density which is eventually negative divergent at the horizon.
If the observation is done at the moment when
the free-fall begins, the energy density is reduced to
\(\epsilon_B(r_s|r_s)=- N[ 4Mr_sf(r_s)+M^2]/[24 \pi
r_s^4 f(r_s)] \), so that the observer who starts  at the horizon
finds the divergent energy density as shown in Fig.~\ref{fig:EB}.

 \subsection{Unruh state}
 \label{sec:unruh}

The Unruh state is characterized by choosing functions of integration as
$t_+=0$ and $t_-=1/(16 \pi M^2)$ in Eq.~\eqref{eq:ed}, which yields the
energy density of
\begin{align}
  \epsilon_U(r|r_s) &= -\frac{N M^2}{12 \pi r^4 f(r)} \Bigg[ \frac{2 r
    f(r_s)}{M} + \frac{f(r_s)}{f(r)} - \frac{1}{2} - \frac{r^4}{64M^4} \left(\sqrt{\frac{f(r_s)}{f(r)}} +
    \sqrt{\frac{f(r_s)}{f(r)}-1} \right)^2 \Bigg]. \label{energy:U}
\end{align}
The free-fall energy density at the horizon
from $r_s$ is simplified as $ \epsilon_U (2M|r_s) = N(63r_s^2-320M
r_s+384M^2 )/[3072 \pi M^2 r_s (r_s-2M )] $,
\begin{figure}[hbt]
  \begin{center}
  \includegraphics[width=0.6\textwidth]{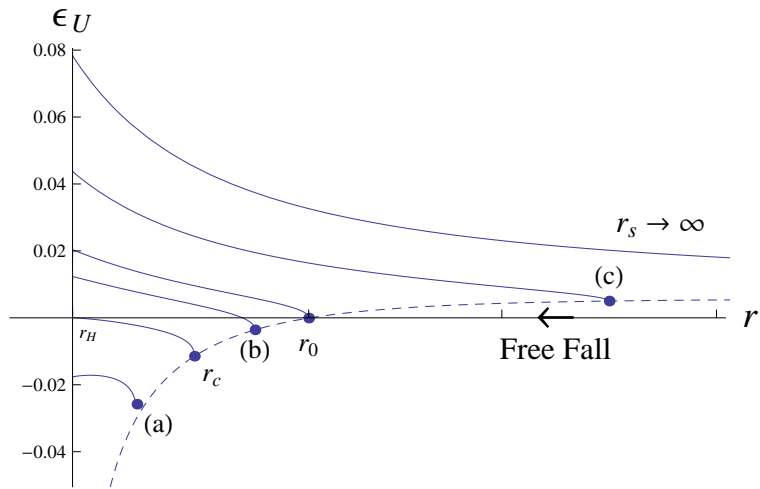}
  \end{center}
  \caption{The energy densities for the Unruh state are plotted by
    setting $N=12,\ M=1$.  The critical point appears at $r_c \approx
    3.1M$ and the energy density vanishes at $r_0 \approx 4.2M$.
    The solid curves are for $ \epsilon_U(r|r_s) $ and the
    dotted curve represents $\epsilon_U (r_s|r_s)$ such that
    there are largely three free-fall cases: (a) is for $r_s < r_c $, (b) is for
    $r_c <r_s<r_0$, and (c) is for $r_s  >r_0$.}
  \label{fig:EUU}  
\end{figure}
which is not always positive definite.
In other words, the initial free-fall position
is crucial to determine the sign of the energy density at the horizon in contrast to
the Boulware case.
Specifically, the energy density at the horizon is indeed positive for
$r_s > r_0$ where $ r_0$ is the initial free-fall position for the energy
density to vanish. For instance, it is positive finite as seen
from the case (c) in  Fig.~\ref{fig:EUU}, and it becomes
$\epsilon_U(2M| \infty) = 21N/(1024 \pi
M^2)$ where the free-fall frame is dropped at the
spatial infinity.
On the other hand, there is a critical point $r_c =
8(20M+\sqrt{22}M)/63$ defined by the point where the observer
finds the zero energy at the horizon, so that the observer would
see the positive energy at the horizon as long as $r_s >r_c$.
 For $r_c < r_s
<r_0$, there appears a transition from the
negative energy density to the positive energy density, which can be seen from the
case (b) in Fig.~\ref{fig:EUU}.
For $r_s <r_c$, the observer will see only negative radiation at the
horizon like the case (a).
When the initial free-fall position approaches the horizon closer,
the larger negative energy density appears.

Using Eq.~\eqref{eq:edr0}, one could obtain the proper energy density at the moment when the
free-fall just begins, where the corresponding energy density is
given as \( \epsilon_U(r_s|r_s) = -NM^2 [ 4 r_s f(r_s)/M + 1 -
r_s^4/(32 M^4) ]/[24 \pi r_s^4 f(r_s)] \) described by the
dotted curve in Fig.~\ref{fig:EUU}.  Note that the energy density
at the horizon $\epsilon_{U}(2M| 2M) $ is
negative divergent, whereas it is positive finite
$\epsilon_{U}(\infty|\infty) \rightarrow \pi (N/12)T_{\rm H}^2$ at the
asymptotic infinity, where $T_{\rm H}$ is the Hawking temperature.

Let us explain why the freely falling observers
moving slowly with respect to the black hole when they pass through the horizon should see very high
(negative) energy density. Actually, the conventional wisdom is that the freely falling observer near the
horizon cannot see any outgoing Hawking radiation as $\langle T_{--}\rangle=0$.
In a collapsing black hole, it was shown that
the energy flow across the future horizon is negative of $\langle T_{++} \rangle <0 $,
 since the corresponding
positive energy would flow out to infinity \cite{Unruh:1977ga}.
Explicitly, the energy density~\eqref{energy}
can be reduced to $\epsilon = \langle T_{++} \rangle u^+u^+ $
at the horizon since $ \langle T_{--}\rangle=\langle T_{+-}\rangle=0$ there.
Note that it does not vanish but also is negative because of non-vanishing ingoing negative flux as
$\langle T_{++} \rangle =-N/(768 \pi M^2) <0 $ from Eq.~\eqref{T:++}. At the horizon, the non-vanishing energy density
is related to the non-vanishing ingoing energy momentum tensor as it should be.
To explain the reason why the high energy density appears near the horizon for a
very slowly falling frame,
let us rewrite the free-fall energy density \eqref{energy}
 as $\epsilon = \langle T_{tt} \rangle u^tu^t$
in the normal coordinates
where the radial velocity is
fixed as $u^{r}=0$ for convenience when the observer is dropped from rest at $r=r_s$.
Note that the time component of the velocity at the stating point of $r_s$ by definition
becomes $u^t=dt/d\tau=1/\sqrt{f(r_s)}$, so that
$dt > d\tau$ where
$dt$ is a time measured by the fiducial observer
 and $d\tau$ is a proper time measured by the freely falling observer.
It shows that the gravitational time dilation effect
is much more significant when the observer is dropped close to the horizon.
As a corollary to this fact, the frequency in the freely falling frame is higher than
that in the fixed frame, so that this factor contributes to the energy density.
Therefore, it becomes the high energy density of $\epsilon(r_s|r_s)  =-N/(768 \pi M^2 f(r_s))$
near the horizon,
where $r_s$ represents the starting position when the observer is dropped from rest.

On the other hand, if the freely falling observer starts with the non-zero initial velocity at
a certain point from the horizon, then the observer can see the positive energy at that instant because
$\langle T_{tr} \rangle$ with $u^r \neq 0$ gives rise to the positive contribution to the energy density.

One more thing to be mentioned is that one could calculate
 the free-fall energy density not only at any finite distance but also near the horizon and at infinity
in the simplified context, which is one of  the advantages of the two-dimensional model,
and one could further discuss
the critical point to characterize the  positive energy zone and the negative energy zone
by solving the exact geodesic equation analytically.
The result shown in Fig.~\ref{fig:EUU} is physically compatible with the previous one that
the positive energy flux would flow out to infinity while a corresponding amount of negative
energy flux would flow down to the black hole \cite{Unruh:1977ga},
so that the area of horizon decreases at a rate
expected positive energy flux at infinity \cite{Candelas:1980zt}.

 \subsection{Hartle-Hawking-Israel state}
 \label{sec:HH}

\begin{figure}[hbt]
  \begin{center}
  \includegraphics[width=0.6\textwidth]{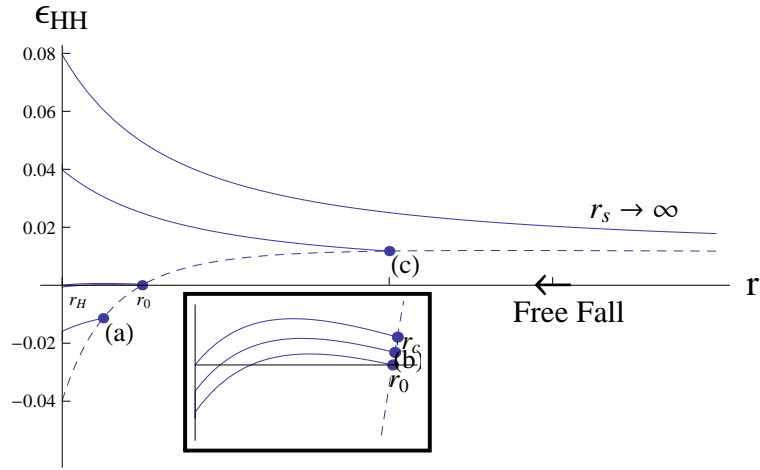}
  \end{center}
  \caption{The energy densities in the Hartle-Hawking-Israel state are plotted
    by setting $N=12,\ M=1$. The sold curves describe $\epsilon_{HH} (r|r_s)$
    and the dotted curve represents $\epsilon_{HH} (r_s|r_s)$.
    There are largely three free-fall cases: (a) $r_s < r_0$, (b)
    $r_0< r_s <r_c$ (box), and (c) $r_s >
    r_c$, where $r_0 \approx 2.98M$  and $r_c=3M$.}
  \label{fig:EHH}  
\end{figure}

For the Hartle-Hawking-Israel state, let us take
$t_\pm=1/(16\pi M^2)$ in Eq.~\eqref{eq:ed}, then the energy density
can be obtained as
\begin{align}
  \label{hh}
  \epsilon_{HH}(r|r_s) = & -\frac{N M^2}{12 \pi r^4 f(r)}
  \left[\frac{2 r f(r_s)}{M}  - \left(\frac{r^4}{16M^4} - 1\right)
    \left(\frac{f(r_s)}{f(r)} - \frac{1}{2}\right)\right].
\end{align}
The freely falling observer at $r_s$ toward the black hole would find the finite
energy density at the horizon of \(\epsilon_{HH}(2M|r_s)=N (r_s - 3M) /
(48\pi M^2 r_s)\). In particular, it becomes
$\epsilon_{HH}(2M|\infty) =
N/(48M^2\pi)$ when the observer is dropped at
spacial infinity with the zero velocity.
There is a point $r_0$ where the proper energy density vanishes; however,
the crucial difference from the Unruh case is that the freely falling
observer starting at $ r_s >r_0$ may encounter alternatively the positive energy and
the negative energy density during the free-fall as shown in the
case (b) in Fig.~\ref{fig:EHH}.
 There is also the critical point $r_c$ to
characterize the sign of the energy density at the horizon,
so that the freely falling observer at the horizon would see the
positive energy density for $r_s >r_c$ and the negative
energy density for $r_s <r_c$ . Moreover,  the
observer would find a transition from the positive
energy density to the negative energy density for $r_0<r_s <r_c$.
Note that $r_0$ and $r_c$ in the Hartle-Hawking-Israel state are smaller than
those in the Unruh state, respectively, which is shown in
Fig.~\ref{fig:EHH}.

 At $r =r_s$, the energy density in the Hartle-Hawking-Israel state
from Eq.~\eqref{eq:edr0} becomes \(
\epsilon_{HH}(r_s|r_s) = - N [8Mr_sf(r_s) + 2M^2 - r_s^4/(8 M^2) ] /
[48 \pi r_s^4 f(r_s)], \) where the behavior of the energy density is
described by the dotted
curve in Fig.~\ref{fig:EHH}.
Explicitly, when $r_s = 2M$, it becomes
negative finite as $\epsilon_{HH}(2M|2M) \rightarrow -N/(96\pi M^2)$
which is contrast to the infinite energy density in the Unruh state.
At the asymptotic infinity, it is finite $
\epsilon_{HH}(\infty|\infty) \to \pi N T_{\rm H}^2/6$, and the energy
density in the Hartle-Hawking-Israel state is two times that of the Unruh
state, $i.e.$, $\epsilon_{HH}(\infty|\infty)=2\epsilon_U(\infty|\infty)$.

\section{Kruskal coordinates at the horizon}
\label{sec:intro3}

From the point of view of an infalling observer,
the gravitational collapse was studied in order to figure out the quantum-mechanical modification of the
collapse \cite{Greenwood:2008ht}.
It was also pointed out that the observers dropped from
a finite distance outside the horizon would detect a finite amount of
radiation when crossing the horizon \cite{Barbado:2011dx, Barbado:2012pt}.
In the exactly soluble two-dimensional Schwarzschild black hole \cite{Eune:2014eka},
the free-fall energy density
was calculated at arbitrary free-fall positions
in order to study the initial free-fall position dependence
of the proper energy density and clarify
whether or not the freely falling observer could encounter
something non-trivial effects at the horizon. In particular,
solving the geodesic equation of motion
over the whole region outside the horizon in the Unruh state,
it could be shown that there exists the negative energy density up to
the extent to the exterior to the horizon of the black hole, roughly $r \sim 3M$
\cite{Eune:2014eka}, where the negative energy zone was introduced in connection with
the firewall argument \cite{Freivogel:2014dca}.
Note that the negative energy density is getting larger and larger
when the initial infalling position from being at rest approaches the horizon
\cite{Eune:2014eka}.
If the observer were dropped at the horizon, actually very near the horizon, the energy-density
 would be divergent \cite{Visser:1996ix,Eune:2014eka,Singh:2014paa}. So,
  one might wonder why the behavior of the energy density at the horizon is different from the conventional result in Ref. \cite{Candelas:1980zt,Birrell:1982ix}.

\subsection{Kruskal coordinates}
\label{Kruskal coordinates}

In order to explain the reason why the different behavior of the energy density appears between
the recent calculation \cite{Eune:2014eka} and the classic works
\cite{Candelas:1980zt,Birrell:1982ix},
let us firstly perform a heuristic calculation by using
the Callan-Giddings-Harvey-Strominger model \cite{Callan:1992rs},
where the length element is given as $ds^2=-e^{2\rho}dx^+dx^-$
with the metric component of
$e^{-2\rho}=M/\lambda-\lambda^2 x^+ x^-$ in the Kruskal coordinates.
The Kruskal coordinates are
related to the tortoise coordinates through the coordinate transformations of
$2\lambda t=\ln(-x^+/x^-)$ and  $2\lambda r^*=\ln(-\lambda^2 x^+ x^-) $,
where $r^*=r+   (1/2\lambda) \ln[1-(M/\lambda) e^{-2\lambda r}]$.
The affine connections in the Kruskal coordinates are calculated as
$\Gamma^+_{++} (x^+,x^-) =2 \partial_+ \rho(x^+,x^-) \sim x^-$,
$\Gamma^-_{--} (x^+,x^-)=2 \partial_- \rho(x^+,x^-) \sim x^+$.
Note that the affine connection of $\Gamma^-_{--} (x^+,0)$ on the future horizon of
$x^-=0$ does not vanish, while
 $\Gamma^+_{++} (x^+,0) =0$. So, the geodesic equation of motion
tells us that $x^-$ cannot be a local flat coordinate on the future horizon,
although the affine connections vanish at $x^{\pm}=0$
corresponding to the bifurcation point.

The awkward situation is not restricted to the above case, and it also
happens in the other models such as the two-dimensional Schwarzschild black hole
 which is actually
of our concern since the model is simple but it  shares most properties in
realistic four-dimensional black holes. The length element
is given as $ds^2=-f(r)dt^2  + f^{-1}(r)dr^2 $ with the metric function of $f(r)=1-2M/r$
in the Schwarzschild coordinates.
The conformal factor for the length element of $ds^2=-e^{2\rho}dx^+dx^-$ in the Kruskal coordinates is obtained as
\begin{equation}
\label{metric2}
e^{2\rho(x^+,x^-)}=\frac{2M}{r}e^{1-\frac{r}{2M}},
\end{equation}
from the conformal transformation of
$x^\pm=\pm 4Me^{\pm \sigma^\pm/4M}$,
where $\sigma^\pm=t\pm r^*$ and
$r^*=r-2M+2M\ln(r/2M-1)$.
The corresponding coordinate transformations
are implemented by
 $t=2M \ln(-x^+/x^-)$ and $r^*=2M\ln(-x^+x^-/(16M^2))$.
At first glance, the affine connections calculated from Eq. \eqref{metric2}  might be expected to vanish
at $r=2M$,
\begin{equation}
\label{ambiguity}
\Gamma^\pm_{\pm\pm} (t,r) =\mp \left(\frac{1}{2r}+\frac{M}{r^2}\right)
\sqrt{\frac{r}{2M}-1}
~~e^{\frac{\mp t-r+2M}{ 4M}}
\end{equation}
in the Kruskal coordinates.
So, it might be tempting to think that the affine connections at $t \rightarrow \infty$
would vanish on the future horizon away from the bifurcation point at a finite $t$ \cite{Candelas:1980zt}.
However, this is not the case.
The two limits such as $r=2M$ and $t \rightarrow \infty$
should be taken at one stroke in order to justify the flatness
on the future horizon,
since the vanishing square root and the divergent exponential function in $\Gamma^-_{--}$
compete on the future horizon.
For this purpose, let us take advantage of
the light cone expressions in the Kruskal coordinates,
then the affine connections \eqref{ambiguity} are neatly calculated as
\begin{align}
\Gamma^+_{++} (x^+,x^-) &= \frac{1}{x^+}\left(\frac{1}{\left(1+W(Z)\right)^2}-1\right), \label{Gppp2} \\
\Gamma^-_{--} (x^+,x^-) &=\frac{1}{x^-}\left(\frac{1}{\left(1+W(Z)\right)^2}-1\right)  \label{Gmmm2},
\end{align}
in virtue of the Lambert $W$ function defined as $Z=W(Z) e^{W(Z)}$ where $Z=-x^+x^-/(16M^2)$.
As a result, the affine connections on the future horizon of $x^-=0$ are written as
\begin{align}
\lim_{x^-\rightarrow0}\Gamma^+_{++} (x^+,x^-)  &=0, \label{limGppp}\\
\lim_{x^-\rightarrow0}\Gamma^-_{--}  (x^+,x^-) &= \frac{x^+}{8M^2} \ne 0, \label{limGmmm}
\end{align}
where  we used the relation of
$W(Z)=Z -Z^2 - O(Z^3)$ near the future horizon.
Note that $\Gamma^-_{--}$
does not vanish on the future horizon, so that it turns out that the coordinate $x^-$ cannot be the
free-fall coordinate.

The above two-dimensional analysis
can be applied to the four-dimensional Schwarzschild black hole whose length element
is given as
$ds^2=-e^{2\rho}dx^+dx^- + r^2 ( d \theta^2 +\sin^2 \theta d \phi^2) $, where
$\rho$ and $r$ are functions of $x^{\pm}$.
The corresponding nonvanishing affine connections on the future horizon are illustrated as
$\Gamma^-_{--}=x^+/(8M^2),
~\Gamma^+_{\theta\theta}=-x^+/2,
~\Gamma^+_{\phi\phi}=(-1/2)x^+ \sin^2\theta,
~\Gamma^\theta_{-\theta}=-x^+/(16M^2),
~\Gamma^\theta_{\phi\phi}=-\cos\theta\sin\theta,
~\Gamma^\phi_{-\phi}=-x^+/(16M^2)$, and
$\Gamma^\phi_{\theta\phi}=\cot\theta$.
We can choose $\theta = \pi/2$ since we are concerned with
the freely falling motion confined on the plane.
In the light of these calculations, the Kruskal coordinates could not be local flat coordinates on the future
horizon except the bifurcation point joining the past horizon and the future horizon corresponding to
$x^{\pm}=0$.
Note that in Ref. \cite{Candelas:1980zt},
the energy-momentum tensors were calculated on the bifurcation two-sphere
for which
$\Gamma^\pm_{\pm\pm} (t,2M)=0$
for any finite time,
and in turn extended to the future horizon by taking infinite time
with a symmetry argument.
If the energy-momentum tensors were calculated directly in the Kruskal coordinates
on the future horizon,
they could not be identified with the energy momentum tensors
in the freely falling frame at that point. So the finiteness of the
energy momentum tensors in Refs. \cite{Candelas:1980zt,Birrell:1982ix} should be reexamined at the future event horizon
in the Unruh state.
\subsection{Energy density in the freely falling frame}
\label{free-fall energy density}

Let us first consider the energy-momentum
tensors in the tortoise coordinates, and
assume that the tensor transformations can be well-defined semiclassically
from the tortoise coordinates to the Kruskal coordinates as a true tensors without
any anomalies such that
\begin{align}
\langle T_{\pm\pm}(x^+,x^-)\rangle &=\left(\frac{\partial \sigma^\pm}{\partial x^\pm} \right)^2 \langle T_{\pm\pm}(\sigma^+,\sigma^-)\rangle,\label{Tpp}\\
\langle T_{+-}(x^+,x^-) \rangle &=\left(\frac{\partial \sigma^+}{\partial x^+} \right)\left(\frac{\partial \sigma^-}{\partial x^-} \right) \langle T_{+-}(\sigma^+,\sigma^-) \rangle.\label{Tpm}
\end{align}
The energy-momentum tensors calculated in the Kruskal coordinates
could not be identified with the proper quantities
except at the bifurcation point as was discussed in the preceding section.
Hence, the right definition for the proper energy density
should be written as
 \begin{equation}
 \label{thanks}
 \epsilon=\langle T_{\tau \tau} \rangle=\frac{d\sigma^{\mu}}{d\tau}  \frac{d\sigma^{\nu}}{d\tau}   \langle T_{\mu\nu}(\sigma^+,\sigma^-) \rangle,
 \end{equation}
where $\tau$ is a proper time.
In other words, the energy-momentum tensors \eqref{Tpp}
and \eqref{Tpm} calculated in the
Kruskal coordinates should be transformed to the local inertial coordinates.
Note that such a form of the energy density \eqref{thanks}
was already introduced  in order to calculate the finite infalling
energy density on the future horizon in
Ref. \cite{Birrell:1982ix}, where
the authors considered an observer moving along a line of constant Kruskal position of $x^1=a$
along with the two-velocity of $(u^{0},~ u^{1})=(dx^0 /d\tau,~dx^1/d\tau)= e^{-\rho}(1,0)$.
The constant spacial radius was expressed in the light cone coordinates as
$ x^+ =x^- +2a$ in the Kruskal coordinates.
However, the constant line does not obey
the geodesic
equation of motion but it can be a geodesic
solution only at the bifurcation point for which $a=0$.
Thus the calculation does not warrant the correctness of the proper energy density
at the future horizon even in spite of
the right definition of the infalling energy density \eqref{thanks}.
In fact, we have repeatedly been asked
why our result is incompatible with the result in
Ref. \cite{Birrell:1982ix}.
Our answer is: ``The proposed geodesic curve proposed in Ref. \cite{Birrell:1982ix} could not actually satisfy the geodesic equation of motion, so that the proper energy density discussed in Ref. \cite{Birrell:1982ix} would not be the correct proper energy density". 

Now, it becomes clear why we have to use
the above definition of the infalling energy density \eqref{thanks} along with
the correct geodesic
solution in order to calculate the energy density
in the freely falling frame.
Using Eq. \eqref{thanks},
we are going to
calculate the infalling energy density on the two-dimensional Schwarzschild black hole
in the Unruh state by means of the light-cone coordinates
in order to avoid any ambiguities.
Let us now start with the conformal gauge fixed energy-momentum tensors \cite{Christensen:1977jc},
\begin{align}
\langle T_{\pm\pm} \rangle &=-\kappa [(\partial_\pm \rho)^2-\partial_\pm^2 \rho+t_\pm], \label{Tpp1}\\
\langle T_{+-} \rangle &= -\kappa \partial_+ \partial_- \rho \label{Tpm1},
\end{align}
which can be derived from the covariant conservation law and the
two-dimensional trace anomaly
for the number  of $N$ massless scalar fields,
and $t_{\pm}$ are the integration functions and $\kappa=N/12 $.
The conformal factor of the two-dimensional Schwarzschild black hole
from Eq. \eqref{metric2}
is written as
\begin{equation}\label{metrictor}
e^{2\rho(\sigma^+,\sigma^-)}=1-\frac{2M}{r(\sigma^+,\sigma^-)},
\end{equation}
in terms of the tortoise coordinates,
 where the radial coordinate is also expressed as
$r(\sigma^+,\sigma^-)=2M(1+W(Y))$ and
by definition $Y=\exp[(\sigma^+-\sigma^-)/4M]$.
From Eqs. \eqref{Tpp1}, \eqref{Tpm1} and \eqref{metrictor},
it is easy to obtain the energy momentum tensors as
\begin{align}
\langle T_{++} \rangle &=-\frac{\kappa}{64M^2}\frac{1+4W(Y)}{(1+W(Y))^4}, \label{Tpp2} \\
\langle T_{--} \rangle &=-\frac{\kappa}{64M^2}\left(\frac{1+4W(Y)}{(1+W(Y))^4}-1\right), \label{Tmm2}\\
\langle T_{+-} \rangle &=\frac{\kappa}{16M^2}\frac{W(Y)}{(1+W(Y))^4}  \label{Tpm2},
\end{align}
which satisfy the Unruh state because we chose $t_+=0$ and $t_-=-1/(64M^2) $
\cite{Unruh:1976db}.
So, the ingoing flux is negative finite on the past horizon from Eq. \eqref{Tpp2},
while
there is no outgoing flux on the future horizon from Eq. \eqref{Tmm2}.
It can be shown that the proper energy density
is finite in the Kruskal coordinate on the future horizon from
the regular coordinate transformation \eqref{Tpm},
which is nothing but the conventional result in Refs.\cite{Candelas:1980zt,Birrell:1982ix}.

As a second step, the components of the two-velocity
are obtained
by exactly solving the geodesic equation of motion for a massive particle as
\begin{equation}\label{upm}
u^\pm (\sigma^+, \sigma^- ; \sigma_s^+, \sigma_s^- )
=\left(\sqrt{1-\frac{1}{1+W(Y_s)}}\pm \sqrt{\frac{1}{1+W(Y)}-\frac{1}{1+W(Y_s)}}\right)^{-1},
\end{equation}
where the initial infalling position at
rest is denoted by $\sigma^{\pm}_s$, and $Y_s=\exp[(\sigma^+_s-\sigma_s^-)/4M]$.
From Eqs. \eqref{Tpp2},\eqref{Tmm2}, \eqref{Tpm2}, and \eqref{upm}, the free-fall energy density \eqref{thanks}
on the future horizon is given as
\begin{equation}\label{density}
\epsilon(\sigma^+, \sigma^- \to \infty ; \sigma_s^+, \sigma_s^- )
 = -\frac{\kappa}{256M^2 W(Y_s)}-\frac{33\kappa}{256M^2}+O(W(Y_s)),
\end{equation}
where the initial infalling position is assumed to be near the future horizon.
It is interesting to note that Eq. \eqref{density} is independent of $\sigma^{+}$, and
just depends on the initial infalling position $\sigma^{\pm}_s$.
Consequently, it turns out that if the observer is dropped extremely on the future horizon for which
$Y_s$ and $W(Y_s)$ vanish, then the proper energy density
is negatively divergent.

\subsection{Blueshift}
\label{blue shift}
We have calculated the proper energy density in the Unruh state
near the future horizon.
Let us now discuss the
origin of the divergence when the observer is dropped
at the horizon as an extreme limit.
Considering a freely falling observer at the initial infalling position of $r_s$
without any journey,
the energy density \eqref{thanks}
is written as
$\epsilon(r_s; r_s) =\langle T_{tt} \rangle u^t u^t +\langle T_{rr} \rangle u^ru^r  +2
\langle T_{tr} \rangle u^t u^r$
in the Schwarzschild coordinates.
When the infalling happens at rest i.e.,  $u^r|_{r_s}=0$,
then the infalling energy density at that moment
is reduced to
$\epsilon(r_s;r_s) =(1/f(r_s))\langle T_{tt} \rangle$
in virtue of $u^{t}|_{r_s}=dt/ d\tau|_{r_s} =1/\sqrt{f(r_s)}$.
Note that the redshift factor is also responsible for
the gravitational time dilation which is larger and larger
as the initial infalling position approaches the horizon.
Next, the value of $\langle T_{tt} \rangle$ in the Schwarzschild coordinates can be
directly obtained
by the use of  the coordinate transformation from
 the tortoise coordinates
to the Schwarzschild coordinates, then the energy density \eqref{thanks} becomes
\begin{equation}
\label{red}
\epsilon(r_s;r_s) =\left.\frac{1}{f(r_s)}[ \langle T_{++} \rangle +
\langle T_{--} \rangle +2 \langle T_{+-}\rangle ] \right|_{r_s},
\end{equation}
where the last term is independent of the vacuum state of black hole and
it can be written as $ \langle T_{+-} \rangle \sim -(\kappa/(16M^2))f(r_s)$ near the horizon.

When the initial infalling position extremely approaches the horizon $r_s \to r_H$,
Eq. \eqref{Tpp1}
can also be expanded asymptotically for each vacuum states.
 First,
the leading order of contributions to the energy-momentum tensors
 in the Boulware state described by choosing $t_{\pm}=0$  \cite{Boulware:1974dm}
becomes finite since $ \langle T_{\pm \pm} \rangle_{\rm{B}} \sim -\kappa/(64M^2)$,
so that the energy density
\eqref{red} is divergent at the horizon.
For the Hartle-Hawking-Israel state implemented by choosing $t_+=t_-=-1/(64M^2) $  \cite{Israel:1976ur,Hartle:1976tp},
the leading order of energy-momentum tensors is written as
$ \langle T_{\pm\pm} \rangle_{\rm{HH}} \sim -(\kappa/(16M^2))f(r_s)$ which
vanish asymptotically at the horizon;
however, the energy density is finite due to the redshift factor
in the denominator in Eq. \eqref{red}.
Hence, these two states result in drastically different conclusions.
By the way, in the Unruh state characterized by $t_+ =0$ and $t_-= -1/(64M^2)$,
the leading order of the energy-momentum tensors near the horizon is calculated as
 $ \langle T_{++} \rangle_{\rm{U}} \sim -\kappa/(64M^2)$ and
$ \langle T_{--} \rangle_{\rm{U}} \sim -(\kappa/(16M^2))f(r_s)$,
where the ingoing flux is negative finite while
the outgoing one vanishes at the horizon.
However, the energy density observed in the freely falling frame
at the horizon is divergent because
the negative finite ingoing flux $ \langle T_{++} \rangle_{\rm{U}} $ is
infinitely blueshifted just like the case of the Boulware state.
Actually, in this case,
the infinite boost is required with respect to the freely falling observer from a finite distance. So the divergent effect is from
moving at the speed of light relative to any infalling
frame that comes from any positive distance outside the horizon.
Thus the divergence is easily explained as a blueshift effect from moving at
the speed of light through radiation.

\section{Effective Tolman Temperature in two dimensions}
\label{sec:intro4}
The proper temperature of a gravitating system for a perfect fluid
in thermodynamic equilibrium has been defined by the well-known Tolman temperature
\cite{Tolman:1930zza, Tolman:1930ona}.
In a static background geometry, it assumes:
(i) the perfect fluid of radiation in thermal equilibrium,
(ii) the covariant conservation law of energy-momentum tensor,
(iii) the traceless condition of energy-momentum tensor,
(iv) the Stefan-Boltzmann law.
The resulting temperature
in the proper frame is written as
\begin{equation}\label{free}
 T_{{\rm T}} = \frac{C}{\sqrt{-g_{00}(r)}},
\end{equation}
where the Tolman factor appears in the denominator and $C$ is a constant determined by a boundary condition.
For example, for the Schwarzschild black hole, the constant could be determined by $C=T_{\rm H}$,
 where $T_{\rm H}$ is the Hawking temperature of the black hole
\cite{Hawking:1974rv,Hawking:1974sw}.
As expected, the Tolman temperature becomes the Hawking temperature at infinity,
whereas it is infinite at the horizon due to the Tolman factor \cite{Frolov:1418196}.
It is worth noting that the Tolman temperature is for the freely falling observer at rest
rather than the fixed observer who undergoes an acceleration \cite{Tolman:1930ona}.
On the other hand, for a fixed observer placed at the radius $r$ of the Schwarzschild black hole,
the temperature
can be expressed as the red/blueshifted Hawking temperature
\begin{equation}\label{fixed}
 T_{{\rm F}} = \frac{T_H}{\sqrt{-g_{00}(r)}},
\end{equation}
where the red/blueshift factor comes from the time dilation
in the presence of the gravitational field at different
places \cite{Weinberg:100595}.
The fixed temperature is infinite at the horizon, which can also be understood in terms of the Unruh effect
for the large black hole
by keeping the detector in place \cite{Unruh:1976db}, since
the Unruh temperature is infinite at the horizon
because of the infinite acceleration of the frame.

Firstly,
it would be interesting to note that the two temperatures \eqref{free} and \eqref{fixed} are
the same in spite of
the complementary observers; the former is for the inertial frame and the latter is for the fixed one.
Secondly, the infinite Tolman temperature
at the horizon is much more puzzling unless $C=0$.
Although the firewall paradox was debated in evaporating black holes \cite{Almheiri:2012rt},
it could also be found even in the static black hole, since
the Tolman temperature \eqref{free} tells us that the freely falling observer encounters
quanta of the super-Planckian frequency at the horizon in the Hartle-Hawking-Israel state \cite{Hartle:1976tp, Israel:1976ur}.
The recent work for the firewall issue in
thermal equilibrium claims the existence of the massless firewall  \cite{Israel:2014eya} whose energy density is negligible but temperature
is infinite at the horizon.
Eventually, it leads to the violation of the equivalence principle at the horizon.

Despite the finite energy density at the horizon in Sec. \ref{sec:ff.frame},
the Tolman temperature is divergent at the horizon.
So, is there any consistent Stefan-Boltzmann law relating the proper energy density to the
temperature? In this section, we will formulate the compatible proper temperature with
the Hawking radiation, and show that the effective Tolman temperature
obtained from the modified Stefan-Boltzmann law is finite everywhere outside
the horizon.

\subsection{Effective Tolman temperature}
\label{sec:Tolman}
We start with a two-dimensional line element given as
\begin{equation}\label{metric3}
ds^2=- f_1(r)dt^2+f_2(r) dr^2,
\end{equation}
where $f_1(r)$ and $f_2(r)$ are static functions and the metric is assumed to be asymptotically flat.
In a static system, the overall macroscopic velocity of radiation flow is zero, and
the velocity can be written as
\begin{equation}\label{velocity1}
u^\mu=\frac{dx^\mu}{d\tau} = \left(\frac{1}{ \sqrt{f_1(r)}},~~0\right).
\end{equation}
The radiation is also regarded as a perfect fluid, so that
the energy-momentum tensor is written as
\begin{equation}\label{Tmunu}
T^{\mu\nu}=(\rho+p)u^\mu u^\nu +p g^{\mu\nu},
\end{equation}
where $\rho=T_{\mu\nu}u^\mu u^\nu$ and $p=T_{\mu \nu} n^\mu n^\nu$ are the local proper energy density and
pressure, respectively, and $n^\mu$ is the spacelike unit normal vector satisfying
$n^\mu n_\mu=1$ and $n^\mu u_\mu=0$.
Note that the flux is also calculated as ${\cal{F}}=-T_{\mu\nu}u^\mu n^\nu$ which is
 zero in the static fluid corresponding to the thermal radiation in equilibrium
\cite{Hartle:1976tp, Israel:1976ur}.
Next, the covariant conservation law of the energy-momentum tensor
can be written as
$2 f_1 \partial_r T^r_r = (T^t_t-T^r_r) \partial_r f_1$,
which is reduced to
\begin{equation}\label{source}
2 f_1 \partial_r p = -(\rho+p) \partial_r f_1.
\end{equation}
Next, the trace equation is given as
\begin{equation}\label{trace}
-\rho + p= T^\mu_\mu,
\end{equation}
where the trace of the energy-momentum tensor is not always zero.
Combining Eqs. \eqref{source} and \eqref{trace}, one can get
\begin{equation}\label{Dfp}
\partial_r (f_1 p)=\frac{1}{2}T^\mu_\mu \partial_r f_1.
\end{equation}
The resulting equation \eqref{Dfp} is easily solved as
\begin{equation}\label{improvedP}
p=\frac{1}{f_1}\left(C_0+\frac{1}{2}\int  T^\mu_\mu  df_1 \right),
\end{equation}
and
\begin{equation}\label{improvedrho}
\rho=\frac{1}{f_1}\left(C_0-f_1 T^\mu_\mu+\frac{1}{2}\int  T^\mu_\mu  df_1 \right),
\end{equation}
where the pressure and energy density are corrected by the non-vanishing trace, respectively.
Here, we will mainly treat the non-trivial trace due to the trace anomaly related to quantum
corrections.

Note that
the conventional Stefan-Boltzmann law in the two dimensional flat space is actually $p=\rho=\alpha T^2$ which is
valid only in the absence of the trace anomaly, where $\alpha$ is the Stefan-Boltzmann constant.
From Eqs. \eqref{improvedP} and \eqref{improvedrho}, the pressure and energy density are no longer
symmetric. To relate the pressure \eqref{improvedP} and energy density \eqref{improvedrho}
to the temperature uniquely, we should find the Stefan-Boltzmann law
which is compatible with the presence of the trace anomaly.

For our purpose, the first law of thermodynamics is written as
\begin{align}
dU =TdS-pdV,
\end{align}
where $U$, $T$, $S$, and $V$ are the thermodynamic internal energy, temperature, entropy, and
volume in the proper frame, respectively, and $U=\int \rho dV$.
Thus, the first law is rewritten in the differential form of
\begin{equation}
\left.\frac{\partial U}{\partial V} \right\vert_T  = T \left.\frac{\partial S}{\partial V}\right\vert_T-p.
\end{equation}
Using the Maxwell relation of $\partial S/\partial V \vert_T = \partial p/\partial T\vert_V$,
we get
\begin{equation}
\label{well}
\rho = T \left.\frac{\partial p}{\partial T}\right\vert_V-p.
\end{equation}
Next, we use the fact that the trace anomaly is independent of the temperature such that $\partial_{T} T^\mu_\mu|_V =0 ~$\cite{BoschiFilho:1991xz},
so that from Eq. \eqref{trace}
we can obtain
\begin{equation}
\label{tt}
\left.\frac{\partial \rho}{\partial T}\right\vert_V =\left.\frac{\partial p}{\partial T}\right\vert_V.
\end{equation}
Plugging Eqs. \eqref{trace} and \eqref{tt} into Eq. \eqref{well} in order to eliminate the pressure and its derivative with
respect to the temperature with the fixed volume,
one can get the first order differential equation for the energy density given as
\begin{equation}
\label{key1}
2\rho = T \left.\frac{\partial \rho}{\partial T}\right\vert_V - T^\mu_\mu.
\end{equation}
Solving Eq. \eqref{key1}, the energy density and pressure can be obtained as
\begin{align}
\label{r}
\rho = \gamma T^2 - \frac{1}{2}T^\mu_\mu,
\end{align}
and
\begin{align}
\label{p}
p = \gamma T^2 + \frac{1}{2}T^\mu_\mu,
\end{align}
where they are reduced to the conventional ones for the traceless case if
the integration constant $\gamma$ is identified with the two-dimensional Stefan-Boltzmann constant, for example, $\gamma=\alpha = \pi/6$
for the massless scalar field \cite{Christensen:1977jc}.
Hence, from Eqs. \eqref{r} and \eqref{p}, the temperature can be written as
\begin{equation}
T=\sqrt{\frac{1}{\alpha}\left(p-\frac{1}{2} T^\mu_\mu \right)} =\sqrt{\frac{1}{\alpha}\left(\rho +\frac{1}{2} T^\mu_\mu \right)}.
\end{equation}
Therefore, the resulting effective Tolman temperature from Eqs. \eqref{improvedP} or \eqref{improvedrho} is obtained as
\begin{equation}\label{improvedT}
T=\frac{1}{\sqrt{\alpha f_1}} \sqrt{C_0 - \frac{f_1}{2}T^\mu_\mu+\frac{1}{2}\int  T^\mu_\mu  df_1 },
\end{equation}
where the temperature is independent of $f_2$.
Indeed, there appeared nontrivial contributions to the temperature from the trace anomaly.
Note that it reduces to the conventional Tolman temperature if the energy-momentum tensor is traceless, so that
$T = C/\sqrt{f_1(r)}$,
where $C =\sqrt{C_0/\alpha}$.
In the asymptotic infinity, the trace parts in Eq. \eqref{improvedT} vanish,
and the constant $C_0$ can be determined by the usual boundary condition.

\subsection{Application to two-dimensional Schwarzschild black hole}
\label{sec:2D Sch}

Let us now show how the effective Tolman temperature
\eqref{improvedT} actually works in the two-dimensional Schwarzschild black hole
described by the metric as
\begin{equation}
\label{metric4}
f(r) = f_1(r) = \frac{1}{f_2(r)} = 1-\frac{2M}{r},
\end{equation}
where $M$ is the mass of black hole and the Newton constant is set to $G=1$.
Now, using the explicit trace anomaly for the massless scalar field as
$T^\mu_\mu = R/(24\pi)$ \cite{Deser:1976yx,Christensen:1977jc},
the proper temperature \eqref{improvedT} can be calculated
by imposing the boundary condition of $C_0 = \alpha/(8\pi M)^2$
which gives the Hawking temperature at infinity,
\begin{align}
\label{tttt}
T=\frac{1}{8\pi M \sqrt{f(r)}} \sqrt{ 1 - 4\left(\frac{2M}{r}\right)^3+3\left(\frac{2M}{r}\right)^4}.
\end{align}
The quantities in the square root in Eq. \eqref{tttt} can be factorized as
\begin{align}
T=&\frac{1}{8\pi M \sqrt{f(r)}} \notag \\
&\times \sqrt{ \left(1 - \frac{2M}{r}\right) \left(1+ \frac{2M}{r}+\left( \frac{2M}{r}\right)^2-3\left( \frac{2M}{r}\right)^3 \right)},
\end{align}
and then the effective Tolman temperature is obtained as
\begin{equation}\label{newT}
T=\frac{1}{8\pi M}\sqrt{1+ \frac{2M}{r}+\left( \frac{2M}{r}\right)^2-3\left( \frac{2M}{r}\right)^3}.
\end{equation}
\begin{figure}[pt]
  \begin{center}
  \includegraphics[width=0.45\textwidth]{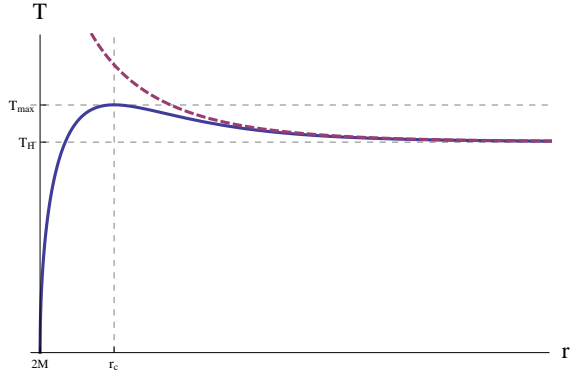}
  \end{center}
  \caption{The thick dotted curve is for the conventional Tolman temperature which is infinite at the horizon, whereas
  the solid curve is for the effective Tolman temperature to be finite
  everywhere.
  The maximum of the latter temperature $T_{\rm max}$
  occurs at $r_c \sim 4M$ in our model. The constant is set to
  $M=1$ for convenience. The infinite Tolman temperature
  at the horizon was suppressed by taking into account the trace anomaly.}
  \label{fig:Tolman}
\end{figure}
Note that the Tolman factor does not appear any more, which is compared to the form of the conventional Tolman temperature
\eqref{free}.
One of the most interesting things to distinguish from the conventional behaviors of the Tolman temperature
is that it is finite everywhere, and it also has a maximum value of
the temperature at $r_c \sim 4M $ as seen from Fig. \ref{fig:Tolman}.
In particular, the temperature vanishes at the horizon.
The suppression of the infinite Tolman temperature at the horizon
with the help of the quantum correction
is reminiscent of the
suppression of the infinite intensity at the high frequency in the black body radiation.

As a matter of fact, for the large black hole, the metric \eqref{metric4}
could be described by the Rindler metric for the near horizon limit.
The Unruh effect tells us that the temperature is given as $T_{\rm U}=a/2\pi$
in terms of the proper acceleration, where the acceleration of the fixed frame is $a= M/( r^2 \sqrt{f(r)})$
\cite{Unruh:1976db}.
It implies that the free-fall observer would find the vanishing Unruh temperature,
if the frame were free from the acceleration.
So, it is reasonable for the observer in the proper frame
to get the vanishing temperature at the horizon rather than
the infinite temperature.
In addition to this, authors  in Ref. \cite{Singleton:2011vh} also showed that
the temperature \eqref{fixed} measured by the fixed observer in
the gravitational background is generically higher than the Unruh temperature of the accelerating observer;
however,
they are the same at the event horizon of the black hole, so that the equivalence principle in the
quantized theory is restored at the horizon.
Thus, the vanishing effective Tolman temperature at the horizon is
compatible with the result that the equivalence principle
is recovered at the horizon \cite{Singleton:2011vh}.

Let us make a comment on the energy density and pressure.
Plugging the effective Tolman temperature \eqref{newT} into Eqs. \eqref{r} and \eqref{p},
one can obtain
\begin{align}
\rho &= - \frac{1}{48\pi r^4 f(r)} \left(8Mr f(r)+2M^2-\frac{r^4}{8M^2}\right) \label{rho1} , \\
p &=\frac{1}{384 \pi M^2} \left[1+ \frac{2M}{r}+\left( \frac{2M}{r}\right)^2+\left( \frac{2M}{r}\right)^3\right] \label{pp},
\end{align}
where the energy density and pressure at the horizon are negative and positive finite as
$-\rho = p=1/(96\pi M^2)$, while $\rho=p = 1/(384\pi M^2)=(\pi/6 )T_H^2$ at infinity.
In a self-contained manner, let us confirm whether the above energy density and pressure calculated by
employing the effective Tolman temperature
are consistent with the results from direct calculations or not.
For this purpose, in the light-cone coordinates defined as $\sigma^{\pm}=t \pm r^*$ through $r^*=r+2M \ln(r/M-2)$,
the proper velocity \eqref{velocity1} can be written as $u^+ =u^- =1 /\sqrt{f(r)}$,
where  $u^\pm=u^t \pm u^r/f(r)$ and
$ n^+ =-n^- =1 /\sqrt{f(r)}$.
The components of the energy-momentum tensor are expressed as
$T_{\pm \pm} = -({1}/{48 \pi}) ({2M
 f(r)}/{r^3} + {M^2}/{r^4} ) + ({1}/{48})t_{\pm}$ and
   $T_{+-}  = -({1}/{48\pi} ) ({2M}f(r)/r^3)$, where $t_{\pm} $ are the integration functions
   obtained from the integration of the covariant conservation law.
The proper energy density and pressure can be
calculated as
$  \rho=  -1/(48 \pi r^4 f(r)) [8Mr f(r)+2M^2 -\pi r^4(t_++t_-) ] $ \cite{Eune:2014eka} and
  $ p=   1/(48 \pi r^4 f(r)) [-2M^2 +\pi r^4(t_++t_-) ]$,
where we used the definition for the free-fall energy density and pressure.
Since the radiation flow in the Hartle-Hawking-Israel state is characterized by choosing the integration functions as
$t_\pm=1/(16\pi M^2)$ \cite{Wipf:1998ss}, one can easily see that
 Eqs. \eqref{rho1} and \eqref{pp} derived from the effective Tolman temperature
\eqref{newT}
are coincident with the proper energy density and pressure from the standard calculations.

It would be interesting to compare our computations with
a previous work. The temperature \eqref{newT} looks different from the free-fall temperature at rest,
  $T_{\rm BT}(r)=(1/8\pi M)$
  $\sqrt{1+ 2M/r+( 2M/r)^2+(2M/r)^3}$ \cite{Brynjolfsson:2008uc}
calculated by using the global embedding of the four-dimensional Schwarzschild black hole into
a higher dimensional flat spacetime \cite{Deser:1998xb}.
For example, the value of $T_{\rm BT}$ at the horizon is larger than that of the temperature at infinity,
precisely, $T_{\rm BT}(2M)=2T_{\rm H}$ which is a maximum.
Simply, we cannot conclude that the difference between them comes from the dimensionality, since
we can exactly get the same free-fall temperature as $T_{\rm BT}$ for
the two-dimensional Schwarzschild black hole \eqref{metric4} by using
an appropriate higher-dimensional embedding method \cite{Banerjee:2010ma}.
Instead, we consider the new expression for the Stefan-Boltzmann law such as
$p=\alpha T^2$ and
$\rho= \alpha T^2 - T^\mu_\mu$, then
$T_{\rm BT}$ can be obtained from Eq. \eqref{pp}; however,
this does not satisfy the thermodynamic relation \eqref{well} which comes from the first law of thermodynamics.
Therefore, if the first law of thermodynamics  is valid in the proper frame,
 the unique Stefan-Boltzmann law can be obtained thermodynamically among diverse expressions to satisfy
the anomaly equation \eqref{trace}.

For the massless firewall in Ref. \cite{Israel:2014eya},
it was claimed that it is massless but hot in the Hartle-Hawking-Israel state of black holes.
At first sight, it seems to be plausible in that
the energy density and pressure at the horizon are at most negligible order of
$1/M^2$ in comparison with
that of the temperature. Moreover, the infinite Tolman temperature at the horizon
indicates the existence of the hot object.
However, employing the effective temperature \eqref{newT},
one could evade the infinite temperature at the horizon, and thus
save the violation of the equivalence principle.

\section{Effective Tolman temperature in four dimensions}
\label{sec:intro5}
 In the Hartle-Hawking-Israel state
 \cite{Hartle:1976tp,Israel:1976ur}, a black hole could be characterized by
 the Hawking temperature $T_{{\rm H}}$ to be proportional to
 the surface gravity. The local temperature in a proper frame
 can be obtained from the blueshifted Hawking temperature of Eq. \eqref{free} as
 \cite{Tolman:1930zza, Tolman:1930ona}
 \begin{equation}\label{localtem}
 T_{{\rm T}} = \frac{T_{\rm H}}{\sqrt{-g_{00}(r)}},
\end{equation}
which is infinite at the horizon due to the infinite blueshift of the Hawking temperature,
though it reduces to the Hawking temperature at infinity.

On the other hand, the renormalized stress tensor
for a conformal scalar field could be finite
 on the background of the
 Schwarzschild black hole \cite{Page:1982fm}.
 At infinity, the proper energy density $\rho$
 is positive finite, which is consistent with
the Stefan-Boltzmann law as $\rho =\sigma T^4_{\rm H}$, where
$\sigma =\pi^2/30$.
If one considered a motion of an inertial
 observer \cite{Page:1982fm, Ford:1993bw, Ford:2009vz}, the negative proper energy density could be found
 near the horizon in various vacua and
its role was also discussed in
 connection with the information loss paradox \cite{Freivogel:2014dca}.
However, it would be interesting
to note that the local temperature \eqref{localtem} is infinite
at the horizon, although the proper energy density
at the horizon $r_{\rm H}$ is negative finite
as $\rho(r_{\rm H}) =-12 \sigma T^4_{\rm H}$ \cite{Page:1982fm}.

 Now, it appears to be puzzling since
 the Tolman temperature at the horizon is positively divergent despite
 the negative finite energy density there.
  More worse, the energy density happens to vanish at a certain point outside the
 horizon \cite{Page:1982fm}, but the local temperature \eqref{localtem} is positive finite at that point.
 In these regards, the Tolman temperature \eqref{localtem} runs contrary to
 the finite renormalized stress tensor, which certainly requires that
 the Stefan-Boltzmann law to relate the stress tensor to the proper temperature
 should be appropriately modified in such a way that they are compatible each other.
This section will be devoted to a generalization of the previous two-dimensional analysis.

\subsection{Proper quantities}
\label{sec:4D}

We start with a four-dimensional
Schwarzschild black hole governed by the static line element as
\begin{equation}
  \label{4Dmetric}
  ds^2 =- f(r)dt^2 + \frac{1}{f(r)} dr^2  + r^2 (d \theta ^2 +\sin^2 \theta d\phi^2),
\end{equation}
where the metric function is $f(r)=1-2GM/r$.
The renormalized stress tensor for a conformal scalar field on
the Schwarzschild black hole was obtained in the Hartle-Hawking-Israel vacuum  \cite{Hartle:1976tp,Israel:1976ur} by using the
Gaussian approximation as~\cite{Page:1982fm}
\begin{align}
  T^\mu_\nu &= \frac{\pi^2}{90} \left(\frac{1}{8\pi M}\right)^4 \left[
              \frac{1-(4-\frac{6M}{r})^2(\frac{2M}{r})^6}{(1-\frac{2M}{r})^2}
              (\delta^\mu_\nu-4\delta^\mu_0\delta^0_\nu) + 24
              \left(\frac{2M}{r}\right)^6 (3\delta^\mu_0\delta^0_\nu +
              \delta^\mu_1\delta^1_\nu) \right], \label{eq:T:ij:Page}
\end{align}
where it is finite everywhere.

On general grounds, the trace anomaly
can be written in the form of curvature invariants as
\begin{equation}
T^\mu_\mu = \alpha \left( \mathcal{F}+ \frac{2}{3} \Box R \right) + \beta \mathcal{G},
\end{equation}
where $\mathcal{F} = R^{\mu\nu\rho\sigma}R_{\mu\nu\rho\sigma} - 2 R^{\mu\nu}
R_{\mu\nu} + R^2/3$ and $ \mathcal{G} = R^{\mu\nu\rho\sigma}R_{\mu\nu\rho\sigma} - 4
R^{\mu\nu} R_{\mu\nu} + R^2$~\cite{Deser:1976yx,
  Duff:1977ay, Birrell:1982ix, Deser:1993yx,Duff:1993wm}.
Actually, there have been
a lot of applications of trace anomalies to Hawking radiation and
black hole thermodynamics in wide variety of cases of interest \cite{Nojiri:1998ue, Nojiri:1998ph, Nojiri:2000ja,
  Elizalde:1999dw, Burinskii:2001bq, Cai:2009ua, Kawai:2014afa,
  Kawai:2015uya, Kawai:2017txu}. The coefficients $\alpha$ and $\beta$ are related
to the number of conformal fields such as real scalar fields
$N_{\rm S}$, Dirac (fermion) fields $N_{\rm F}$, and vector fields
$N_{\rm V}$, such that they are fixed as
$\alpha = (120 (4\pi)^2)^{-1} (N_{\rm S} + 6 N_{\rm F} + 12 N_{\rm
  V})$ and $\beta = - (360 (4\pi)^2)^{-1} (N_{\rm S} + 11 N_{\rm F} + 62 N_{\rm
  V})$. For the Ricci flat spacetime with a single conformal scalar field, the trace
anomaly reduces to
\begin{align}
  T^\mu_\mu = \frac{1}{2880\pi^2}R^{\mu\nu\rho\sigma}
  R_{\mu\nu\rho\sigma}= \frac{M^2}{60\pi^2 r^6}, \label{eq:4D:Sch:anomaly}
\end{align}
and the trace for the stress tensor~\eqref{eq:T:ij:Page} is exactly
in accord with the conformal
anomaly~\eqref{eq:4D:Sch:anomaly}.

In contrast to the two dimensional case~\cite{Gim:2015era},
the stress tensor appears anisotropic in the spherically symmetric black
hole in four dimensions, and so the form of the stress
tensor~\eqref{eq:T:ij:Page} should be generically written as~\cite{Hawking:1973,
  Lobo:2007zb}
\begin{align}
  T^{\mu\nu} = (\rho+p_t)u^\mu u^\nu +p_t g^{\mu\nu}+(p_r-p_t)
  n_{(r)}^\mu n_{(r)}^\nu.   \label{eq:EM:anistropic}
\end{align}
The proper velocity $u^\mu$ is a timelike unit vector satisfying
$u^\mu u_\mu = -1$, $n^\mu_{(r)}$ is the unit spacelike vector in
the radial direction, and $n_{(\theta)}^\mu$ and $n_{(\phi)}^\mu$
are the unit normal vectors which are orthogonal to $n^\mu_{(r)}$
satisfying $g_{\mu \nu} n_{(i)}^\mu n_{(j)}^\mu = \delta_{ij}$ and
$n_{(i)}^\mu u_\mu=0$ where $i,j = r,\theta, \phi$.  Thus the
spacelike unit normal vectors are determined as
\begin{align}
  n^\mu_{(r)} &=\left(0, \sqrt{f(r)}, 0, 0
                \right), \quad n^\mu_{(\theta)}=\left(0, 0, \frac{1}{r}, 0 \right),
                \quad n^\mu_{(\phi)}=\left( 0, 0, 0,
                \frac{1}{r \sin\theta} \right), \label{n}
\end{align}
with the proper velocity
\begin{align}
  u^\mu = \left(\frac{1}{ \sqrt{f(r)}}, 0, 0, 0\right)   \label{4Dvelocity}
\end{align}
for the frame dropped from rest.  Then, from
Eqs.~\eqref{eq:T:ij:Page}, \eqref{eq:EM:anistropic}, \eqref{n},
and~\eqref{4Dvelocity}, the proper energy density and pressures can be
explicitly calculated by using the following relations,
\begin{align}
  \rho = T_{\mu\nu} u^\mu u^\nu,~p_r = T_{\mu \nu} n_{(r)}^\mu
  n_{(r)}^\nu,~ p_t = T_{\mu \nu} n_{(\theta)}^\mu n_{(\theta)}^\nu =
  T_{\mu \nu} n_{(\phi)}^\mu n_{(\phi)}^\nu, \label{proper}
\end{align}
where the proper flux along $x^i$-direction can also be obtained by using
the relation $\mathcal{F}_i=-T_{\mu\nu}u^\mu n_{(i)}^\nu$ but it
trivially vanishes in thermal equilibrium~\cite{Hartle:1976tp, Israel:1976ur}.

Note that the energy density and pressures are not independent as seen from the
trace relation,
\begin{align}
  T^\mu_\mu = -\rho + p_r + 2 p_t.   \label{4Dtrace}
\end{align}
From Eqs.~\eqref{eq:T:ij:Page}, \eqref{eq:4D:Sch:anomaly},
and~\eqref{proper}, we find an additional relation
\begin{align}
  p_r-p_t = \frac{1}{4} T^\mu_\mu , \label{eq:p.r:p.t}
\end{align}
which characterizes the anisotropy
between the tangential pressure and radial pressure.

Let us now express the proper energy density and pressures formally in
terms of the trace anomaly for our purpose.  From
Eq.~\eqref{eq:EM:anistropic}, the covariant
conservation law for the energy-momentum tensor is rewritten as
\begin{align}
  \partial_r p_r + \frac{2}{r} (p_r-p_t) + \frac{1}{2f} \partial_r f
  (p_r+\rho)  =0.   \label{4Dsource}
\end{align}
Plugging Eqs.~\eqref{4Dtrace} and~\eqref{eq:p.r:p.t}
 into Eq.~\eqref{4Dsource}, one can obtain the simplified form of
\begin{align}
  \partial_r  p_r + \frac{\partial_r f}{2f}p_r = - \left(\frac{1}{2r}
    + \frac{3 \partial_r f}{4f}\right)T^\mu_\mu,   \label{4DDfp}
\end{align}
which can be solved as
\begin{align}
  p_r = \frac{1}{f^2} \left(C_0 + \int^r \frac{f}{4r}(-2f
  +3r \partial_r f)  T^\mu_\mu  dr \right),   \label{4DimprovedPr}
\end{align}
where $C_0$ is an integration constant.
Additionally, from Eqs.~\eqref{4Dtrace} and \eqref{eq:p.r:p.t}, the
tangential pressure and energy density can also be obtained as
\begin{align}
  p_t &= \frac{1}{f^2}\left(C_0-\frac{f^2}{4}T^\mu_\mu+\int^r
        \frac{f}{4r}(-2f +3r \partial_r f)  T^\mu_\mu  dr
        \right),\label{4DimprovedPt} \\
  \rho &= \frac{3}{f^2}\left(C_0-\frac{f^2}{2}T^\mu_\mu+\int^r
         \frac{f}{4r}(-2f +3r \partial_r f)  T^\mu_\mu  dr \right). \label{4Dimprovedrho}
\end{align}
The above proper quantities
were nicely related to the trace anomaly.
 \subsection{Effective Tolman temperature from anisotropic fluid}
\label{sec:4}

We derive the proper temperature for the background
of the four-dimensional Schwarzschild black hole by using
the modified Stefan-Boltzmann law.
First of all, we note that the volume of the system in the radial proper frame can
be changed only along the radial direction on the spherically
symmetric black hole,
and the thermodynamic first law can be written
as
\begin{align}
  dU = T dS - p_r dV  \label{eq:1st.law}
\end{align}
without recourse to the tangential work. From Eq.~\eqref{eq:1st.law}, we
can immediately get
\begin{align}
  \left(\frac{\partial U}{\partial V} \right)_T = T
  \left(\frac{\partial S}{\partial V} \right)_T -
  p_r.   \label{eq:1st.law:another}
\end{align}
Next, from the Maxwell relations such as
$(\partial S/\partial V)_T= (\partial p_r /\partial T)_V$,
we obtain
\begin{align}
  \rho =  T \left(\frac{\partial p_r}{\partial T} \right)_V -
  p_r.   \label{eq:eom:therm.}
\end{align}
Using the fact that the trace anomaly is independent of temperature as
$\partial_{T} T^\mu_\mu =0$~\cite{BoschiFilho:1991xz},
from Eqs.~\eqref{4Dtrace} and \eqref{eq:p.r:p.t}, we also obtain
\begin{align}
  \left(\frac{\partial \rho}{\partial T}\right)_V
  =\left(\frac{\partial p_r}{\partial
      T}\right)_V+2\left(\frac{\partial p_t}{\partial
      T}\right)_V, \label{partial p1}
\end{align}
and
\begin{equation}
  \left(\frac{\partial p_r}{\partial T}\right)_V=\left(\frac{\partial
      p_t}{\partial T}\right)_V. \label{partial p2}
\end{equation}
Plugging Eqs.~\eqref{partial p1} and \eqref{partial p2} into
Eq.~\eqref{eq:eom:therm.}, we get
\begin{equation}
 \label{eq:4D:p.r:eq}
 T \left(\frac{\partial \rho}{\partial T}\right)_V-4\rho = \frac{3}{2}T^\mu_\mu,
\end{equation}
which is solved as
\begin{align}
  \rho = 3\gamma T^4 - \frac{3}{8}T^\mu_\mu.\label{4Dr}
\end{align}
From Eqs.~\eqref{4Dtrace} and~\eqref{eq:p.r:p.t}, the radial and
tangential pressure are also derived as
\begin{align}
  p_r &= \gamma T^4 + \frac{3}{8}T^\mu_\mu, \label{4Dpr} \\
  p_t &= \gamma T^4 + \frac{1}{8}T^\mu_\mu, \label{4Dpt}
\end{align}
respectively.  The integration constant $\gamma$ is related to
the Stefan-Boltzmann constant $\sigma$ as $\gamma = \sigma/3=\pi^2/90$~ for a
conformal scalar field~\cite{Christensen:1977jc}.
For the traceless case,
the modified Stefan-Boltzmann law \eqref{4Dr}
simply reduces to the usual one.  The proper energy density in Eq.~\eqref{4Dr} is not
necessarily positive definite thanks to the trace anomaly, so that
the negative energy states are naturally permitted in this extended setting.

From Eqs.~\eqref{4Dr}, \eqref{4Dpr}, and \eqref{4Dpt}, the proper
temperature is obtained as
\begin{align}
  T = \left[\frac{1}{\gamma} \left(p_r-\frac{3}{8} T^\mu_\mu \right)
  \right]^{1/4}  = \left[\frac{1}{\gamma} \left(p_t-\frac{1}{8}
  T^\mu_\mu \right) \right]^{1/4}  = \left[\frac{1}{3\gamma}
  \left(\rho+\frac{3}{8} T^\mu_\mu \right) \right]^{1/4},
\end{align}
which can be compactly written in terms of the trace anomaly as
\begin{equation}\label{eq:4D:T}
  T =\frac{1}{\gamma^{1/4} \sqrt{f}} \left( C_0 - \frac{3}{8} f^2
    T^\mu_\mu+\int^r \frac{f}{4r}(-2f +3r \partial_r f)  T^\mu_\mu  dr
  \right)^{1/4},
\end{equation}
where we used Eqs.~\eqref{4DimprovedPr}, \eqref{4DimprovedPt}, and
\eqref{4Dimprovedrho}.  In the absence of the trace anomaly, the
proper temperature~\eqref{eq:4D:T} reduces to the usual Tolman
temperature~\cite{Tolman:1930zza, Tolman:1930ona}.
Requiring that the proper temperature~\eqref{eq:4D:T} be coincident with
the Hawking temperature $T_{\rm H}$ at infinity, we can fix the constant
as $C_0=\gamma^{1/4} T_{\rm H}$.

\begin{figure}[hpt]
  \begin{center}
  \includegraphics[width=0.7\linewidth]{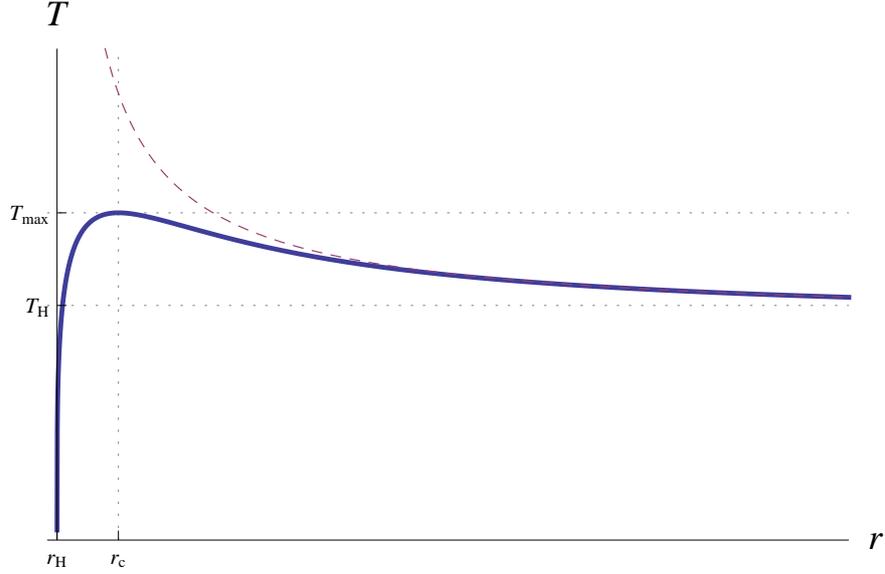}
  \end{center}
  \caption{The dashed curve shows the behavior of the usual Tolman temperature
    of being divergent on the horizon. The solid curve is
    the effective Tolman temperature, which is
    finite everywhere. In particular, it vanishes at the horizon and has
    a maximum $T_{\rm max} \sim 1.51 T_{\rm H}$ at
    $r_c \sim 1.31 r_H$. All the curves approach the Hawking
    temperature at infinity, whereas they are very different from each
    other near the horizon where quantum effects are significant.}
  \label{fig:Tvsr}
\end{figure}

Finally, plugging the trace
anomaly~\eqref{eq:4D:Sch:anomaly} into Eq.~\eqref{eq:4D:T}, we obtain
\begin{align}
  T=\frac{1}{8\pi M \sqrt{f(r)}} \left[ 1 -
  28 \left(\frac{2M}{r}\right)^6 + 48 \left(\frac{2M}{r}\right)^7 - 21
  \left(\frac{2M}{r} \right)^8 \right]^{1/4},
\end{align}
which can be neatly factorized as
\begin{align}
  T = \frac{1}{8\pi M \sqrt{f(r)}}
  &\left[\left(1 - \frac{2M}{r}\right)^2 \left(1 +
    2\left(\frac{2M}{r}\right) + 3\left(\frac{2M}{r}\right)^2
    \right. \right. \notag \\
  & \left. \left. + 4 \left(\frac{2M}{r}\right)^3 +
    5\left( \frac{2M}{r}\right)^4 + 6\left( \frac{2M}{r}\right)^5 -
    21\left( \frac{2M}{r}\right)^6 \right) \right]^{1/4}.
\end{align}
It would be interesting to note
that the blueshift factor in the denominator related to the origin of the divergence at the horizon
can be canceled out, so that the
effective Tolman temperature is written as
\begin{align}
  T = \frac{1}{8\pi M}\left[\left(1 - \frac{2M}{r} \right) \sum_{n=1}^6
  \frac{n(n+1)}{2} \left(\frac{2M}{r}\right)^{n-1} \right]^{1/4}. \label{4DnewT}
\end{align}
Thus the redshift factor responsible for the
infinite blueshift of the Hawking temperature on the horizon does not
appear any more in the effective Tolman temperature.
As seen from Fig.~\ref{fig:Tvsr}, the behavior of the temperature~\eqref{4DnewT}
shows that it is finite everywhere and approaches the Hawking
temperature at infinity.  In particular, it is vanishing on the
horizon, so that the freely falling observer from rest does not see any excited particles.
On the contrary to the naively expected divergence from the
usual Tolman temperature at the horizon, the high frequency quanta
could not be found on the horizon, which would be compatible with the result that
the equivalence principle could be recovered at the
horizon~\cite{Singleton:2011vh}.

The divergent
dashed curve near the horizon in Fig.~\ref{fig:Tvsr} could be made finite
by taking into account the quantum effect via the trace anomaly, which is reminiscent of the
vanishing Hawking temperature in the noncommutative Schwarzschild black
hole based on the different assumptions of quantization rules~\cite{Nicolini:2008aj}.
And the proper temperature based on the effective temperature method
is also compatible with the present result in the sense that
the proper temperature vanishes at the horizon \cite{Barbado:2016nfy}.

\section{Origin of Hawking radiation and firewall}
\label{sec:intro6}

Here, we elucidate how the Hawking radiation and the firewall appear
simultaneously in a tractable field theoretic model,
and then provide a compelling argument for their compatibility between
the firewall and the Hawking radiation.
The key is to decompose quantitatively the Tolman temperature
read off from the Stefan-Boltzmann law
into the two chiral temperatures of $T_\text{L}$, $T_\text{R}$
defined by the negative influx and the positive outward flux, respectively, and then figure out
their properties carefully.
It will be shown that $T_\text{L}$ becomes infinite at the horizon, which is regarded as a signal of the
firewall; however, it vanishes at infinity, so that it does not affect the asymptotic observer at infinity.
The essential reason for the existence of the firewall is
due to the infinitely blueshifted negative influx crossing the horizon
rather than the outward flux.
On the other hand, $T_\text{R}$ will be shown to be finite everywhere by identifying it
with the effective Tolman temperature presented in the preceding sections.
In particular, it vanishes at the horizon and approaches the Hawking temperature
at infinity. Thus, it shows that the
outgoing Hawking radiation originates from the atmosphere of the near-horizon quantum region,
not just at the horizon.
After all, the present analysis in the semiclassically fixed background approximation will show
that the firewall is not only a definite physical object but also a natural solution in the Unruh vacuum,
and the Hawking radiation indeed originates from the atmosphere without any conflicts with the presence of the firewall.

\subsection{Energy density and flux}
\label{sec:Energy density and flux}

Let us start with a two-dimensional general static black hole described by the metric,
\begin{eqnarray}
ds^2=- f(r)dt^2+\frac{1}{f(r)}dr^2,
\end{eqnarray}
where $f(r)$ is an asymptotically flat metric function. The constants are set to $\hbar=k_{\text B}= G=c=1$.
The event horizon $r_\text{H}$ is defined by $f(r_\text{H})$=0, and
the Hawking temperature is calculated from the definition of the surface gravity as
$T_\text{H}=f'(r_\text{H})/4\pi$  \cite{Hawking:1974sw}
where the prime denotes the derivative with respect to $r$.
The Hawking temperature is blueshifted
for a distant observer outside the horizon \cite{Wald:1999xu}, which is simply
written as the Tolman form \cite{Tolman:1930zza}.

From the covariant conservation law and the conformal anomaly of $\langle T^\mu_\mu \rangle= R/(24\pi)$
for a two-dimensional massless scalar field
\cite{Deser:1976yx}, the components of the stress tensor are determined as
$\langle T_{\pm\pm} \rangle =\left(ff''-({1}/{2})f'^2 +t_\pm  \right)/{(96\pi)}$,
$\langle T_{+-} \rangle =ff''/{(96\pi)}$,
where $t_\pm$ reflect the non-locality of the conformal anomaly \cite{Christensen:1977jc}.
Note that the expectation value of the energy-momentum tensor was
written by using the tortoise coordinates $\sigma^{\pm}=t \pm r^{*}(r)$ where $r^*=\int dr/f(r)$,
but $f(r)$ was just written in terms of $r$ instead of $r^*$ for convenience \cite{Giddings:2015uzr}.
Then, the proper energy density, pressure, and flux can be defined as
$\rho=\langle T_{\mu\nu}\rangle u^\mu u^\nu $, $p=\langle T_{\mu \nu} \rangle n^\mu n^\nu$, and ${\cal{F}}=-\langle T_{\mu\nu} \rangle u^\mu n^\nu$,
where $u^\mu$ is a two-velocity and $n^\mu$ is a spacelike unit normal vector satisfying
$n^\mu n_\mu=1$ and $n^\mu u_\mu=0$.
Explicitly, in the light-cone coordinates,
the velocity vector from the geodesic equation of motion
and the normal vector are solved in a freely falling frame from rest as \cite{Eune:2014eka}
\begin{equation}
u^+ =u^- = n^+ =- n^- =\frac{1}{\sqrt{f}}. \label{n1}
\end{equation}
In particular,
the proper energy density and flux are expressed by
\begin{eqnarray}
\rho &=& \frac{1}{f} \left( \langle T_{++} \rangle +\langle T_{--} \rangle + 2 \langle T_{+-} \rangle \right) \label{rho}, \\
{\cal{F}} &=& -\frac{1}{f} \left( \langle T_{++} \rangle -\langle T_{--} \rangle  \right), \label{flux}
\end{eqnarray}
where the pressure is related to the energy density via
the trace relation of $\langle T^\mu_\mu \rangle=-\rho + p$.
From Eqs. \eqref{rho} and \eqref{flux}, the energy density and flux are explicitly written
as $\rho = \left(4ff''-f'^2 +t_+ +t_- \right)/(96\pi f)$ and
${\cal{F}}  = - \left( t_+ - t_- \right)/(96\pi f)$.

In the Hartle-Hawking-Israel vacuum \cite{Hartle:1976tp,Israel:1976ur},
the stress tensor is regular at both the future horizon and the past horizon,
so that the regularity condition determines the
integration constants as $t_+=t_-=(1/2)f'^2(r_\text{H})$.
Let us assume that the metric function is finite at least up to the second derivative with $f'' <0$, for
instance, which holds for the Schwarzschild black hole or the CGHS black hole \cite{Callan:1992rs},
then the curvature scalar of $R=-f''$ is positive finite.
The proper energy density \eqref{rho} also becomes finite everywhere. In particular,
it is negative finite at the horizon, $\rho_\text{HH}(r_\text{H})=f''(r_\text{H})/(48\pi)$,
while it is positive finite at
infinity, $\rho_\text{HH}(\infty)= f'^2(r_\text{H})/(96\pi)$. It shows that the proper energy density is not always positive.

There is another equilibrium state defined by $t_+=t_-=0$
called the Boulware vacuum \cite{Boulware:1974dm}.
The energy density \eqref{rho} is negatively divergent at the horizon, $\rho_{\text B}(r_\text{H}) \to -\infty$
and negatively vanishes at infinity, $\rho_{\text B}(\infty) =0$.
If such a black hole exists, then it will be surrounded by the negative energy density in equilibrium.
Note that the energy density is divergent at the horizon,
so that the smoothness of the horizon is not warranted.

The Unruh vacuum of our interest is defined
by twisting two equilibrium states asymmetrically
in such a way that $t_+ =0$ and $t_- =f'^2(r_\text{H})/2$ which describes an evaporating black hole
semiclassically \cite{Unruh:1976db}.
Then the proper energy density is negative infinity
at the horizon, $\rho_{\text U}(r_\text{H}) \to -\infty$ like the case of the Boulware vacuum \cite{Boulware:1974dm}.
It is positive finite at infinity like the case of the Hartle-Hawking-Israel vacuum
defined by $t_+=t_-=(1/2)f'^2(r_\text{H})$ \cite{Hartle:1976tp,Israel:1976ur}, but its magnitude is half
of that of the Hartle-Hawking-Israel vacuum, i.e., $\rho_\text{U}(\infty)= f'^2(r_\text{H})/(192\pi)=\rho_\text{HH}(\infty)/2$.
As expected, the non-vanishing flux is obtained as ${\cal{F}}_\text{U}(r)= f'^2(r)/(192\pi f)$ which
is coincident with the energy density at infinity, but it is
divergent at the horizon.

Note that the above expectation value
of the energy-momentum tensor $\langle T_{\mu\nu} \rangle$  in any vacua is finite at the horizon,
whereas the proper flux and the energy density in the Unruh vacuum and the energy density in
the Boulware vacuum are divergent there.
One might wonder what the origin of the divergence is.
The expectation value of the energy-momentum tensor
$\langle T_{ab}(\xi^a,\xi^b) \rangle$ defined in a locally inertial coordinate system
can be obtained from the general coordinate transformation of
$ \langle T_{\mu\nu} \rangle$ defined in the tortoise coordinate system of $\sigma^{\pm}=t \pm r^{*}(r)$,
which is implemented by
\begin{equation}
\langle T_{ab} \rangle=\frac{\partial \sigma^\mu}{\partial \xi^a}
\frac{\partial \sigma^\nu}{\partial \xi^b} \langle T_{\mu\nu} \rangle,
\label{tensor}
\end{equation}
where $a,b=0,1$ are the indices for the locally inertial coordinate system and
$\mu,\nu=\pm$ are for the tortoise coordinate system.
We also consider the coordinate transformation of the velocity vector and the unit normal spacelike vector as
\begin{equation}
u^\mu = \frac{\partial \sigma^\mu}{ \partial \xi^a } u^a=\frac{\partial \sigma^\mu}{ \partial \xi^0 },~~~~~~
n^\mu = \frac{\partial \sigma^\mu}{ \partial \xi^a } n^a=\frac{\partial \sigma^\mu}{ \partial \xi^1 } \label{velocity}
\end{equation}
where $u^a=(1,0), n^a=(0,1)$ are defined in the locally inertial coordinate system.
Therefore, the flux  from Eq. \eqref{tensor} is written as
\begin{equation}
{\cal{F}}=-\langle T_{01} \rangle=-\frac{\partial \sigma^\mu}{\partial \xi^0} \frac{\partial
\sigma^\nu}{\partial \xi^1} \langle T_{\mu\nu} \rangle=-u^\mu n^\nu
\langle T_{\mu\nu} \rangle, \label{t01}
\end{equation}
by using Eq. \eqref{velocity}, which is nothing but Eq. \eqref{flux} when Eq. \eqref{n1} is used.
The energy density \eqref{rho} can also be obtained from the coordinate transformation of
$\langle T_{00} \rangle=(\partial \sigma^\mu /\partial \xi^0) (\partial \sigma^\nu /\partial \xi^0)
\langle T_{\mu\nu} \rangle=u^\mu u^\nu
\langle T_{\mu\nu} \rangle$, which reproduces Eq. \eqref{rho}.
As a result, in the Unruh vacuum, the divergent proper quantities
at the horizon comes from the singular coordinate transformation
at the horizon.

\subsection{Stefan-Boltzmann law in equilibrium}
\label{sec:Stefan-Boltzmann law in equilibrium}

Conventionally, the Stefan-Boltzmann law in thermal equilibrium rests upon
the traceless condition of the stress tensor~\cite{Tolman:1930zza};
however, it is worth noting that the trace anomaly is responsible for the
Hawking radiation~\cite{Christensen:1977jc}.
In that sense, assuming the nontrivial trace of the stress tensor,
one should obtain a modified Stefan-Boltzmann law which gives an effective Tolman temperature
induced by the trace anomaly \cite{Gim:2015era}.
Let us now obtain the effective Tolman temperature from the modified Stefan-Boltzmann law
in the Hartle-Hawking-Israel vacuum prior to the discussion in the case of the Unruh vacuum,
and then write
the effective temperature in terms of a much more convenient form for our purpose.

From Eq.~\eqref{r}, the energy density is related to the effective Tolman temperature as
\begin{align}
 \rho= \gamma T^2_\text{eff} -\frac{1}{2} \langle T^\mu_\mu \rangle  \label{effenergy}
\end{align}
where  $\gamma = \pi/6$
for a massless scalar field~\cite{Christensen:1977jc}.
The modified Stefan-Boltzmann law \eqref{effenergy}
simply reduces to the usual Stefan-Boltzmann law of $\rho=\gamma T^2$
for the traceless case, which yields the usual Tolman temperature.

Note that
the proper energy density in the Hartle-Hawking-Israel vacuum is not always positive \cite{Visser:1996ix}.
Thus, one can find
the position $r_0$ where the energy density vanishes,
so that the region of the positive energy density is separated from that of the negative energy density.
For example, the position is estimated as $r_{0} \sim 2.98 M$
in the two-dimensional Schwarzschild black hole \cite{Eune:2014eka}.
It should be noted that
the usual Stefan-Boltzmann law holds only at infinity such as $\rho_\text{HH}(\infty)=\gamma T_{\text H}^2$
; however, it appears to be
unreliable in
the region of the negative energy density for $r_{\text H} <r<r_0$.
Fortunately, the energy density need not
be positive thanks to the anomalous term in the modified Stefan-Boltzmann law \eqref{effenergy}.

Now, plugging the expression for the energy density in Eq.~\eqref{rho} into Eq.~\eqref{effenergy},
one can get the effective Tolman temperature in the Hartle-Hawking-Israel vacuum as
\begin{eqnarray}
\label{eff}
\gamma T^2_\text{eff} =\frac{1}{f} \left( \langle T_{++} \rangle_{\text{HH}} + \langle T_{--} \rangle_{\text{HH}} \right)
 = \frac{1}{96\pi f}\left( 2ff''-f'^2 +f'^2(r_\text{H}) \right),
\end{eqnarray}
where the stress tensors are calculated with respect to the Hartle-Hawking-Israel vacuum.
In contrast to the divergent usual Tolman temperature at the horizon,
the effective Tolman temperature \eqref{eff} vanishes at the horizon,
which can be shown by taking the limit at the horizon.
In fact, there is neither influx nor outward flux at the horizon, and so there is no reason for the firewall to exist.
This fact is consistent with the regularity of the renormalized stress tensor at the horizon
in the Hartle-Hawking-Israel vacuum and
the explicit calculation to make use of the detector \cite{Singleton:2011vh}.

\subsection{Stefan-Boltzmann law in a radiating system}
\label{sb}

Let us derive the temperature for a radiating system such as the black hole in the Unruh vacuum.
The temperature is obtained from
the radiated power which is just the proper flux. The proper temperature
in the Unruh vacuum can be read off from
the two-dimensional Stefan-Boltzmann law \cite{Giddings:2015uzr},
\begin{equation}
\label{Giddings}
\sigma T^2 =-\frac{1}{f} \left(\langle T_{++} \rangle_{\text U} -\langle T_{--} \rangle_{\text U} \right)
=\frac{\pi}{12}\left(  \frac{T_\text{H} }{ \sqrt{f}} \right)^2,
\end{equation}
where the stress tensors are calculated with respect to the Unruh vacuum.
The two-dimensional Stefan-Boltzmann constant in a radiating system $\sigma$
is half of that of the equilibrium state, so that $\sigma=\gamma /2=\pi/12$ is consistent.
Note that the ingoing and outgoing stress tensors in the Unruh vacuum are
proportional to those in the Boulware and Hartle-Hawking-Israel vacua, respectively.
So the black hole temperature in the Unruh vacuum \eqref{Giddings} can be decomposed into
the left and right chiral temperatures
without mixing chirality as
\begin{eqnarray}
\sigma T_\text{L}^2 =-\frac{\langle T_{++}\rangle_{\text{B}}   } {f},~~~~~~  \sigma T_\text{R}^2 =\frac{\langle T_{--} \rangle
_{\text{HH}}}{f}, \label{leftright}
\end{eqnarray}
where we used the relations of
$\langle T_{++}\rangle_{\text{U}}=\langle T_{++}\rangle_{\text{B}}$ and
$\langle T_{--} \rangle_{\text{U}}=\langle T_{--} \rangle_{\text{HH}}$.

Equilibrium states of black holes are commonly described by the  Boulware and Hartle-Hawking-Israel vacua.
This is possible only when the systems are locally in equilibrium and sufficiently slowly varying.
 In contrast to these vacua,
 the net flux is not zero for the Unruh vacuum,
  so that the black hole in this state is not in equilibrium as seen from Eq. \eqref{Giddings}.
   If the two equilibrium systems are interacting,
   then the thermal temperature for a better interpretation will require a quasi-equilibrium condition
    between the two different equilibrium systems. However, in the present semiclassical approximation,
    the influx and outward flux are actually decoupled and they do not interfere.

For the left temperature in Eq. \eqref{leftright}, the ingoing flux is negative finite at
the horizon as $\langle T_{++}(r_{\text H}) \rangle_{\text B}=-f'^2(r_\text{H}) /(192\pi)$,
so that the left temperature becomes positively divergent at the horizon as $T_\text{L} (r_H) \rightarrow +\infty$.
So the firewall in the Unruh vacuum arises from
the infinite blueshift of the negative ingoing flux despite the absence of the outgoing flux at the horizon,
so that the Tolman temperature \eqref{Giddings} is divergent at the horizon.
However, the ingoing flux decreases to zero and does not
reach infinity, so that $T_\text{L}(\infty)=0$.
It implies that the ingoing superplanckian excitations have no impact on the
asymptotic observer at infinity. Thus, these excitations are certainly responsible for the firewall
but completely decoupled from the Hawking radiation at infinity.

Next, after rescaling $\sigma  =\gamma /2$, the right temperature in Eq. \eqref{leftright}
can be shown to be equivalent to the effective Tolman temperature \eqref{eff},
\begin{align}
T_\text{R}=T_\text{eff}, \label{good}
\end{align}
by use of the equilibrium condition of $\langle T_{++}\rangle_{\text{HH}}=\langle T_{--}\rangle_{\text{HH}} $.
Thus, the right temperature
directly possesses the same properties as those of the effective Tolman temperature,
so that it is finite everywhere.
In particular, it vanishes at the horizon and approaches the Hawking temperature exactly
at the asymptotic infinity.
Therefore, it shows that the Hawking particles at infinity
originate from the atmosphere outside the horizon
rather than the firewall.

On the other hand, from the left temperature in Eq. \eqref{leftright},
it would be interesting to note that one can also define a black hole temperature
in the Boulware vacuum in a manner similar to the way of the Hartle-Hawking-Israel vacuum
by replacing $\sigma  = \gamma /2$ and
$\langle T_{++}\rangle_{\text{B}} =\left(\langle T_{++}\rangle_{\text{B}} +\langle T_{--}\rangle_{\text{B}}\right)/2 $
since $\langle T_{++}\rangle_{\text{B}}=\langle T_{--}\rangle_{\text{B}} $.
So, the left-right temperatures in Eq.~\eqref{leftright} can be compactly written in the unified manner as \cite{Kim:2016iyf}
\begin{align}
\label{wonder}
\gamma T^2_\text{L,R} =\mp \frac{1}{f} \left( \langle T_{++} \rangle_{\text{B,HH}} + \langle T_{--} \rangle_{\text{B,HH}} \right),
\end{align}
where the local Boulware temperature and the local Hawking temperature
are eventually on an equal footing, and thus, the ingoing particles are also in thermal states like the outgoing particles.
It implies that the particles in the same chirality can be entangled with their partners,
but the particles in a different chirality need not be entangled since they are not
created from pair creation \cite{Israel:2015ava}.
Importantly, this would be one of the advantages of the present analysis without recourse to pair creation,
which could respect the monogamy principle in quantum mechanics.
\begin{figure}[pt]
  \begin{center}
  \includegraphics[width=0.45\textwidth]{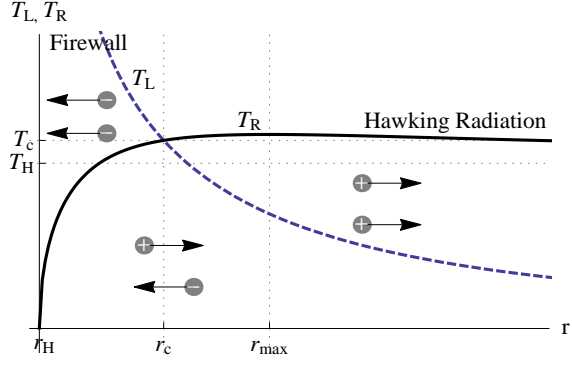}
  \end{center}
  \caption{The dashed curve is for $T_{\text L}$, and
   the solid one is for $T_{\text R}$, where $M=1$ for simplicity.
   The minus-plus signs in the small circles with the left and right arrows mean the negative influx
   and the positive outward flux, respectively.
      The critical position of the flux transition occurs at
    $r_{\text c} \sim 3.26M$ and the corresponding critical temperature $T_{\text c }$ is
   slightly higher than the value of the Hawking temperature $T_{\text H}$, where
   $T_c /T_{\text H} \sim 1.14$. The maximum of the right
   temperature occurs at $r_{\text{max}} \sim 4.32M$.
   }
  \label{fig:1}
\end{figure}

The origin of the Hawking radiation and
the reason for the existence of the firewall have been discussed based on the generic metric.
Let us now discuss the arguments explicitly
for the two-dimensional Schwarzschild black hole
described by $f(r)=1- 2M/r$.
From Eq. \eqref{wonder}, one can obtain the left and right temperatures
as \cite{Kim:2016iyf}
\begin{align}
{\rm Firewall}: T_{\rm L}&=\frac{1}{\sqrt{2}\pi r} \sqrt{\frac{M}{r-2M}\left(1-\frac{3M}{2r}\right)}, \label{L}\\
{\rm Hawking~~radiation}:T_{\rm R}&=\frac{1}{8\pi M} \sqrt{1+\frac{2M}{r}+\left(\frac{2M}{r}\right)^2-3\left(\frac{2M}{r}\right)^3}. \label{R}
\end{align}
Note that the left temperature is found to be infinite at the horizon
and vanishes at infinity, which means that the firewall appears significantly at the horizon and
the asymptotic observer is free from the impact of
the firewall. (See also a recent application of the left temperature in Ref. \cite{Good:2016atu}.)
On the other hand, the right temperature is finite everywhere,
which shows that the outgoing radiation is a very low energy phenomenon since it is almost comparable
to the Hawking temperature over the entire region outside the horizon.
In particular, it vanishes at the horizon and approaches the Hawking temperature at infinity.
As a corollary, from Eqs. \eqref{L} and \eqref{R}, the Tolman temperature obtained as $T= \sqrt{ T_\text{L}^2 + T_\text{R}^2}$
shows only the collective behavior of the two chiral temperatures.

In Fig. \ref{fig:1}, one may define a critical position $r_c$ at which
the two excitation energies are the same as $T_{\text L}=T_{\text R}$. It yields $r_{\text c} \sim 3.26M$ which
is a larger macroscopic distance as compared to the horizon size.
For $r <r_{\text c}$, the left temperature is dominant and
eventually predicts the firewall at the horizon, whereas
the right temperature is dominant for $r >r_{\text c}$ and reproduces the Hawking temperature at infinity.
Moreover, $T_{\text R}$ has a peak at $r_{\text{max}} \sim 4.32M$
which is larger than $r_{\text c}$. Thus, the critical transition from the influx to the outward flux occurs
before the right temperature arrives at the peak.

\section{The initial radiation energy density in warm inflation}
\label{sec:intro7}

One of the most important ingredients
in the thermal history of the universe is
to determine the temperature at the end of inflationary regime, $i.e.$, the reheating temperature, $T_{\rm r}$, in the standard inflation models.
Even though the exact value of the reheating temperature has not yet been known,
the upper bound of the temperature has been estimated as the scale of the grand unified theory (GUT),
 $T_{\rm r} \leq 10^{16}~{\rm GeV}$ \cite{Kolb1990},
 and the lower bound of the temperature has been constrained by the big bang nucleosynthesis as
 $T_{\rm r} \geq 10~{\rm Mev}$ \cite{Kawasaki:2000en, Hannestad:2004px}.
 Afterwards, another lower bound of the reheating temperature has been derived from the CMB data
based on the seven year Wilkinson
microwave anisotropies probe (WMAP7) data as
$T_{\rm r} \geq 6~{\rm Tev}$ \cite{Martin:2010kz}.
On the other hand, in the warm inflation model, the order of the temperature $T_{\rm end}$ at the end of inflation was obtained
as $T_{\rm end} \sim 10^{13} ~{\rm GeV}$ \cite{Hall:2003zp}.
In the presence of the non-minimal kinetic coupling model,
the temperature at the end of inflation was calculated
up to the uncertainties of the cosmological observations
as $5.01\times 10^{7} {\rm GeV} \le T_{\rm end} \le 2.11\times 10^{13}{\rm GeV}$
by use of the formalism introduced in Ref. \cite{Mielczarek:2010ag}
with the data of Planck 2013
\cite{Goodarzi:2014fna}.

In the warm inflation scenario,
the radiation is closely in thermal equilibrium,
and thus
the initial radiation density is naturally assumed to be nonzero, $\rho_{\rm r}(t_{\rm i}) \neq 0$ \cite{Berera:2008ar},
which is
compatible with the Stefan-Boltzmann law of $\rho_r =3 \gamma T^4$
in the hot thermal bath at the initial point of inflation, $t=t_{\rm i}$.
On the other hand,
it was claimed that
the initial radiation energy density in thermal equilibrium with the thermal bath
is unjustified, and so it is required that $\rho_{\rm r}(t_{\rm i}) = 0$ when inflation starts,
following the spirit of the chaotic inflation scenario that the universe
should be created form a quantum fluctuation of the vacuum \cite{Bellini:2001ka}.
In this new scenario, the initial zero temperature is also increasing during inflation
based on the framework of the warm inflation scenario. Now, one might wonder
how to get the non-vanishing radiation energy density in warm inflation scenario.
In connection with
the non-zero initial radiation energy density in warm inflation scenario,
thermodynamic analysis is given for the warm inflation model
by using the definitions for the inflaton and radiation energy density presented in Ref. \cite{Kolb1990}.

From the modified Stefan-Boltzmann law,
it will be shown that the zero radiation energy density at
the Grand Unification epoch just prior to starting inflation becomes finite when inflation starts,
which will give a sufficient radiation energy density after inflation.
By using the number of e-folds and the spectral index of the scalar perturbation under the slow-roll approximations in the power-law potential and damping terms,
it will be found that the temperature at the end of warm inflation successfully
gives the upper bound lower than the GUT scale \cite{Kolb1990},
and lower bound of the big bang nucleosynthesis~\cite{Kawasaki:2000en, Hannestad:2004px}
in the regime of the CMB data \cite{Martin:2010kz}.
Consequently, the sufficient radiation energy density is produced at the end of inflation.

\subsection{Modified Stefan-Boltzmann law in warm inflation}
\label{sec:warminf}

One of the most important ingredients in warm inflation is that the decreasing radiation energy density during inflation
is replenished in such a way that the energy of the inflaton field is transferred to that of radiation in virtue of dissipation.
It is worth noticing that only the radiation energy density is related to the temperature via
the Stefan-Boltzmann law in the standard warm inflation models.
As compared to this,
if one were to treat the inflaton and radiation on an equal footing in equilibrium,
then one might encounter generically non-vanishing trace of the total energy-momentum tensor due to the inflaton part
while the radiation part is still traceless.
Now, it should be emphasized that the usual
Stefan-Boltzmann law commonly rests upon the traceless condition of the energy-momentum tensor,
and thus we have to modify the Stefan-Boltzmann law in order
to incorporate the non-vanishing trace of the total energy-momentum tensor.

Let us start with the Helmholtz free energy defined by
$F = E-T S$, where $E$, $T$, and $S$ are the energy, temperature, and entropy
of a thermal system, respectively.
From the differential form of the Helmholtz free energy as $dF=dE-TdS-SdT$,
one can obtain the relation between the entropy $S$ and the Helmholtz free energy as $S=-\partial F/ \partial T$.
Using the Euler's relation of $E=TS-p {\rm V}$, one can also rewrite the Helmholtz free energy as
$F=-p {\rm V}$ where $p$ is the pressure and ${\rm V} $ is the volume of the system.
Then, it yields a relation for the entropy density
of $s= S/{\rm V}$ as
\begin{equation}\label{eq:spT}
s=\frac{\partial p}{\partial T}.
\end{equation}
On the other hand, the relevant energy-momentum tensor is assumed to be perfect fluid written as
$ T_{\mu\nu}=(\rho+p)u_\mu u_\nu +  g_{\mu\nu} p $,
where $u^\mu$ is the four-velocity of radiation flow satisfying $u^\mu u_\mu = -1$.
Assuming that the trace of the energy-momentum tensor is non-vanishing generically,
the trace relation is obtained as
$-\rho + 3p= T^\mu_\mu$ with the Euler's relation rewritten as
$ \rho +p=Ts$ where $\rho = E/{\rm V}$.
By eliminating the pressure in Eq. \eqref{eq:spT}, the differential equation for the energy density is obtained as
\begin{align}
T\frac{\partial \rho}{\partial T}-4 \rho &=T^\mu_\mu-T \frac{\partial T^\mu_\mu}{\partial T},
\end{align}
so that the modified Stefan-Boltzmann law to incorporate the non-vanishing trace of the energy-momentum tensor
can be obtained as
\begin{align}
\rho(T)=3 C_0 T^4 - \frac{1}{4}T^\mu_\mu - \frac{3}{4}T^4 \int^T \frac{1}{T^4}\frac{\partial T^\mu_\mu }{\partial T}dT,  \label{eq:SBrho}
\end{align}
and the pressure of
\begin{align}
p(T)= C_0 T^4 + \frac{1}{4}T^\mu_\mu - \frac{1}{4}T^4 \int^{T} \frac{1}{T^4}\frac{\partial T^\mu_\mu }{\partial T}dT, \label{eq:SBp}
\end{align}
where the integration constant $C_0$ can be fixed from an initial condition.
The relations \eqref{eq:SBrho} and \eqref{eq:SBp} naturally reduce  to the usual Stefan-Boltzmann law for the traceless case,
so that $C_0 =\gamma$, where $\gamma$ is the Stefan-Boltzmann constant.
However, $C_0$ will be fixed for the case of the non-vanishing trace for our purpose later
by imposing a different boundary condition.
In fact, such a modified Stefan-Boltzmann law induced from conformal anomalies
had been applied to SU(3) lattice gauge theory
in particle physics in the Minkowski spacetime \cite{Boyd:1996bx} and the recent black hole physics
in connection with the information loss problem \cite{Gim:2015era}.

From the cosmological point of view, let us assume that the total system of the early universe
consists of inflaton and radiation in thermal equilibrium.
Then  the total energy density $\rho_{\rm tot}$ and pressure $p_{\rm tot}$ are written
as \cite{Kolb1990}
\begin{align}
\rho_{\rm tot} &=\rho_\phi+\rho_{\rm r}= \frac{1}{2}\dot{\phi}^2 + V_{\rm eff}(\phi,T) +\rho_{\rm r}, \label{eq:rhotot} \\
p_{\rm tot} &=p_\phi+p_{\rm r} = \frac{1}{2}\dot{\phi}^2 - V_{\rm eff}(\phi,T) + p_{\rm r}, \label{eq:ptot}
\end{align}
where $\rho_{\rm r},~p_{\rm r}$ and $\rho_{\phi},~p_\phi$ denote the energy density and pressure of radiation
and inflaton, respectively. Specifically, the temperature dependent
effective potential $V_{\rm eff}$  for the inflaton is expressed by
\cite{Dolan:1973qd, Weinberg:1974hy, Linde:1978px}
\begin{equation}\label{eq:potential}
V_{\rm eff}(\phi, T)= - \gamma T^4 + \frac{1}{2}(\delta m_T)^2 \phi^2+V_0(\phi),
\end{equation}
where $\gamma=\pi^2 g_{*} /90$ and $g_{*}$ is an effective particle number.
$V_0(\phi)$ is the zero-temperature potential for the scalar field $\phi$, and $\delta m_T(\phi,T)$ denotes a thermal correction which will be neglected for simplicity along the lines of Ref. \cite{Hall:2003zp}.

The traceless condition for the radiation leads to
the equation of state as $p_{\rm r}=(1/3)\rho_{\rm r}$; however,
the trace for the total energy-momentum tensor
appears non-vanishing due to the effective potential for the inflaton as
\begin{align}
T^\mu_\mu =-\rho_{\rm tot}+ 3p_{\rm tot}  =  - 4V_{\rm eff}(\phi,T), \label{eq:trace}
\end{align}
where the kinetic energy is assumed to be very small as compared to the potential energy from now on.
By plugging Eq. \eqref{eq:trace} into Eqs. \eqref{eq:SBrho} and \eqref{eq:SBp}, the explicit forms of the pressure and energy density are obtained as
\begin{align}
\rho_{\rm tot} &= 12\gamma T^4 \ln \left( \frac{T_0}{T} \right) - \gamma T^4 +V_0(\phi), \label{eq:rhotot2}  \\
p_{\rm tot} &= 4\gamma T^4 \ln \left( \frac{T_0}{T} \right)+\gamma T^4 -V_0(\phi) \label{eq:ptot2}
\end{align}
by using the initial condition of $C_0= 4\gamma \ln T_0$
from the assumption that there exists only the inflaton field at the initial temperature of our universe $T_0$,
$i.e.$, $\rho_{{\rm tot}}(T_0)=\rho_\phi$ and $p_{{\rm tot}}(T_0)=p_\phi$.
Now, we take $T_0$  to be the GUT temperature as the maximum temperature of our universe
$T_0=T_{\rm GUT} = 10^{16} {\rm GeV}$, since
all perturbative interactions can be frozen out
and ineffective in maintaining or establishing thermal equilibrium for $T > 10^{16} {\rm GeV}$, and thus
the known interactions are not capable of thermalizing the universe at temperature greater than the GUT scale \cite{Kolb1990}.
Thus the energy density \eqref{eq:rhotot2} and pressure \eqref{eq:ptot2} are written as
\begin{align}
\rho_{\rm tot} = V_{\rm eff}+3\gamma T^4 \ln \left( \frac{T_{\rm GUT}}{T} \right)^4,~~
p_{\rm tot} =  -V_{\rm eff}+\gamma T^4 \ln \left( \frac{T_{\rm GUT}}{T} \right)^4. \label{eq:ptot3}
\end{align}
Comparing Eq. \eqref{eq:ptot3} with Eqs. \eqref{eq:rhotot} and \eqref{eq:ptot}, we can immediately find
the modified Stefan-Boltzmann law for the radiation as \cite{Gim:2016uvv}
\begin{align}
\rho_{\rm r} &= 3 \gamma T^4 \ln \left( \frac{T_{\rm GUT}}{T} \right)^4,\qquad
p_{\rm r} = \gamma T^4 \ln \left( \frac{T_{\rm GUT}}{T} \right)^4.   \label{eq:radrhop}
\end{align}
The traceless condition for the radiation is still met as $\rho_{\rm r} = 3p_{\rm r}$
which has been used not only in the warm inflation scenario \cite{Berera:1995ie, Hall:2003zp} but also
in the variety of cases of interest, for example, in the warm inflation model with the non minimal kinetic coupling \cite{Goodarzi:2014fna} and tachyon warm inflationary model \cite{Herrera:2006ck}.

It is interesting to note that
the energy density and pressure \eqref{eq:radrhop} for the radiation could be formally expressed
as the usual Stefan-Boltzmann law of $\rho_{\rm r} = 3 \gamma_{\rm eff} T^4$ and $p_{\rm r} = \gamma_{\rm eff} T^4$
when defining the temperature-dependent Stefan-Boltzmann constant
as $\gamma_{\rm eff}(T)=\gamma \ln \left( T_{\rm GUT}/T \right)^4$. So, the non-vanishing trace of the energy-momentum
tensor for the inflaton is of relevance to the modification of the Stefan-Boltzmann constant.
The physical consequence of this modification is that the radiation energy density vanishes at $T_{\rm GUT}$
and then it increases when the temperature of the universe decreases.
Subsequently, when inflation starts, it becomes finite
and gives the adequate energy density for radiation, and finally it
reaches $\rho_{\rm end}$ at $T_{\rm end}$ at the end of inflation.
This fact provides the reason why the original model for warm inflation could assume
the finite radiation energy density when inflation starts.
 In essence, the radiation energy density could be thermodynamically created before inflation starts.

One might wonder why the form of the present Stefan-Boltzmann law
\eqref{eq:radrhop} is different from the previous one in Ref. \cite{Hall:2003zp}.
Apart from the additional consideration of the non-vanishing total energy-momentum tensor \eqref{eq:trace}
in the present thermodynamic analysis,
the other reason would stem from the different treatment of the temperature dependent term in
the finite temperature effective potential for the inflaton of $V_{\rm eff}(\phi, T)$.
From the total energy density \eqref{eq:rhotot} and  pressure \eqref{eq:ptot} with the effective potential \eqref{eq:potential}, the energy densities  and the pressures were identified with $\tilde{\rho}_{\phi}=\dot{\phi}^2/2 + V_0,~\tilde{p}_{\phi}=\dot{\phi}^2/2 - V_0$ and $\tilde{\rho}_{\rm r}=-\gamma T^4+\rho_{\rm r},~\tilde{p}_{\rm r}=\gamma T^4+p_{\rm r}$ in Ref. \cite{Hall:2003zp}.
In this case, the temperature-dependent term of $-\gamma T^4$ in $V_{\rm eff}(\phi, T)$ was incorporated into the radiation energy density and pressure rather than the inflaton energy density and pressure, so that the usual forms were obtained as
$\tilde{\rho}_{\rm r}=3 \gamma T^4,~\tilde{p}_{\rm r}=\gamma T^4$
by using the solution of $p_{\rm r}=0,~\rho_{\rm r} =4\gamma T^4$ obtained
from $\tilde{\rho}_{\rm r}+\tilde{p}_{\rm r}=T \tilde{s}_{\rm r},~\tilde{\rho}_{\rm r}=3\tilde{p}_{\rm r}$,
where $\tilde{s}_{\rm r}=\partial \tilde{p}_{\rm r}/\partial T$.
However, there is another way to identify the pressure such as
 $p_\phi = -V_{\rm eff}(\phi,T)$ with the effective potential \eqref{eq:potential} \cite{Kolb1990}.
In this case,
the temperature dependent term in $V_{\rm eff}(\phi, T)$ was not included in the radiation part but it was incorporated in
the inflation part,
since it was originated from loop-corrections of inflaton field in thermal bath.
For the latter choice, Eq. \eqref{eq:radrhop} could be obtained \cite{Gim:2016uvv}.

\subsection{Slow-roll approximations}
\label{sec:SR}
We consider the inflaton interacting with the radiation,
and thus the equations describing the system show how the energy lost by the inflaton through the damping force
is transferred to the radiation.
In the warm inflation model \cite{Hall:2003zp}, the conservation law, $\dot{\rho}_{\rm tot}+3H(\rho_{\rm tot}+p_{\rm tot})=0$,
can be separated into the inflaton and radiation parts as
\begin{align}
\dot{\rho_\phi}+3H(\rho_\phi+p_\phi) &= -\Gamma \dot{\phi}(t)^2, \label{eq:EOM1} \\
\dot{\rho_{\rm r}}+3H(\rho_{\rm r}+p_{\rm r}) &= \Gamma \dot{\phi}(t)^2,  \label{eq:EOM2}
\end{align}
where $H=\dot{a}/a$ denotes the Hubble parameter, and
$\Gamma \dot{\phi}^2$ is the friction term adopted phenomenologically
to describe the decay of the inflaton field and its energy transfers into the radiation bath.
And, the Friedmann equation for the evolution of the universe is also given as
\begin{equation}\label{eq:EOM3}
H^2-\frac{1}{3m_{\rm p}^2}\rho_{\rm tot}=0.
\end{equation}

Now, we exhibit slow-roll approximations to neglect terms of the highest order in time derivatives
in Eqs. \eqref{eq:EOM1}, \eqref{eq:EOM2}, and \eqref{eq:EOM3}
with the assumption that the inflaton field is dominant over the radiation field during the slow-roll warm inflation \cite{Hall:2003zp}. So, we obtain
\begin{equation}\label{eq:SRapprox1}
\dot{\phi}^2 \ll V_{\rm eff} , \qquad \ddot{\phi} \ll \Gamma \dot{\phi} , \qquad
\dot{\rho_{\rm r}} \ll 4H \rho_{\rm r}, \qquad \rho_{\rm r} \ll \rho_\phi.
\end{equation}
By using a set of slow-roll parameters,
\begin{align}
\epsilon = \frac{m_{\rm p}^2}{2}\left( \frac{\partial_\phi V_{\rm eff}}{V_{\rm eff}}\right), \qquad \eta = m_{\rm p}^2\left( \frac{\partial_\phi^2 V_{\rm eff}}{V_{\rm eff}}\right),
\qquad \beta = m_{\rm p}^2\left( \frac{\partial_\phi V_{\rm eff}}{V_{\rm eff}}\right)\left( \frac{\partial_\phi \Gamma}{\Gamma} \right), \label{eq:SRparameters}
\end{align}
 the slow-roll approximations \eqref{eq:SRapprox1} can be summarized
as $\epsilon \ll r,~~\eta \ll r,~~ \beta \ll r $,
where $r$ is the ratio of the production rate of radiation, $\Gamma$, to the expansion rate, $3H$, defined as $r \equiv \Gamma/(3H)$.
Note that the slow-roll conditions are applied to the finite temperature effective potential \eqref{eq:potential}
rather than $V_0(\phi)$ which is contrast to the slow-roll conditions employed
in the standard warm inflation \cite{Berera:1995ie, Hall:2003zp}.
Neglecting several terms in Eqs. \eqref{eq:EOM1}, \eqref{eq:EOM2}, and \eqref{eq:EOM3}
within the slow-roll approximations \eqref{eq:SRapprox1}, one can get the following reduced equations,
\begin{align}
 3H r \dot{\phi} + \partial_\phi V_{\rm eff} &= 0, \label{eq:SReq1} \\
 3H (\rho_{\rm r}+p_{\rm r}) -\Gamma(\phi) \dot{\phi}^2 &= 0, \label{eq:SReq2} \\
H^2-\frac{1}{3m_{\rm p}^2}V_{\rm eff} &= 0,  \label{eq:SReq3}
\end{align}
where $r \gg 1$ in the warm inflationary regime.
Combining Eqs. \eqref{eq:SReq1} and \eqref{eq:SReq3}, one can rewrite Eq. \eqref{eq:SReq2} as
$\rho_{\rm r}+p_{\rm r} = m_{\rm p}( \partial_\phi V_{\rm eff})^2 /(\sqrt{3V_{\rm eff} }\Gamma)$,
and then obtain
\begin{equation}\label{eq:DiffT}
 4\gamma T^4 \ln \left( \frac{T_{\rm GUT}}{T} \right)^4  = \frac{m_{\rm p}( \partial_\phi V_{\rm eff})^2}{\sqrt{3V_{\rm eff}} \Gamma}
\end{equation}
by using the expressions for the energy density and pressure \eqref{eq:radrhop}.

Next, the number of e-folds during warm inflation is given as
\begin{align}
N_{\rm inf}=\int^{t_{\rm end}}_{t_{\rm HC}} H(t) dt
=\int^{\phi_{\rm HC}}_{\phi_{\rm end}} \frac{\Gamma \sqrt{V_{\rm eff}}}{\sqrt{3}m_{\rm p} \partial_\phi V_{\rm eff}}d\phi ,\label{eq:efold}
\end{align}
where $\phi_{\rm HC}$ and $\phi_{\rm end}$ are the values of the inflaton field corresponding  to
the horizon-crossing time $t_{\rm HC}$ and the end time of warm inflation $t_{\rm end}$, respectively.
Hereafter, in order to perform the specific calculations, we adopt the power-law potential $V_0$
and damping term $\Gamma$  \cite{Hall:2003zp} as
\begin{equation}\label{eq:Ex}
V_0(\phi)=\lambda \phi^n,  \qquad
\Gamma(\phi)=\Gamma_0 \left( \frac{\phi}{\phi_0}\right)^m,
\end{equation}
where the coefficients $\Gamma_0,~\phi_0$ and $\lambda$ are constants,
 and the power $n$ and $m$ are fixed as $n=2,~m=2$ for simplicity.
In this specific model, the number of e-folds \eqref{eq:efold} during inflation era is
finally written as
\begin{align}
N_{\rm inf} = \frac{\Gamma_0 (\lambda \phi_{\rm HC}^2-\gamma_{\rm HC} T^4_{\rm HC})^{\frac{3}{2}}}{6\sqrt{3} m_{\rm p} \lambda^2 \phi_0^2}  \label{eq:NSR}
\end{align}
by assuming that $\phi_{\rm end} \ll \phi_{\rm HC}$.

\subsection{Cosmological perturbations and temperature}
\label{sec:perturb}
Let us determine the temperature bound at the end of inflation via cosmological perturbation.
The thermal fluctuations produce the power spectrum $P_\zeta$ for the comoving curvature $\zeta$ \cite{Hall:2003zp},
\begin{equation}\label{eq:P1}
P_\zeta =  \frac{\pi^{\frac{1}{2}}H^{\frac{5}{2}}\Gamma^{\frac{1}{2}}T}{2\dot{\phi}^2},
\end{equation}
and the power spectral index $n_{\rm s}$ for the scalar perturbation
is defined as $n_{\rm s}-1 = d\ln |P_\zeta |/d\ln k$,
which is calculated as
\begin{align}
n_{\rm s}-1 =\frac{5}{2H}\frac{d\ln H}{dt}+\frac{1}{2H}\frac{d\ln \Gamma}{dt}-\frac{2}{H}\frac{d\ln \dot{\phi}}{dt}+\frac{1}{H}\frac{d\ln T}{dt} \label{eq:ns1}
\end{align}
at the horizon crossing defined as $k=aH$,
where the relation $d\ln k \approx d\ln a = Hdt$ was employed.
Using the slow-roll equations \eqref{eq:SReq1}, \eqref{eq:SReq2}, and \eqref{eq:SReq3} with the entropy density,
we obtain the relations
\begin{equation}\label{eq:SRparameter2}
\frac{1}{H}\frac{d\ln H}{dt} = -\frac{1}{r} \epsilon, \qquad
\frac{1}{H}\frac{d\ln \Gamma}{dt} = -\frac{1}{r}\beta, \qquad
\frac{1}{H}\frac{d\ln \dot{\phi}}{dt} = \frac{1}{r}(\beta -\eta),
\end{equation}
and
\begin{equation}\label{eq:SRparameter3}
\frac{1}{H}\frac{d\ln T}{dt} = \frac{1}{4r}\left(1+\frac{1}{\ln\left(\frac{T_{\rm GUT}}{T}\right)^4-1} \right)(\epsilon+\beta-2\eta).
\end{equation}
From Eqs. \eqref{eq:SRparameter2} and \eqref{eq:SRparameter3},
we can express the spectral index in terms of slow-roll parameters \eqref{eq:SRparameters} as
\begin{equation}\label{eq:ns2}
n_{\rm s}-1 = \frac{3\eta}{2r}-\frac{9}{4r}(\epsilon+\beta)+\frac{1}{4r}\left(1+\frac{1}{\ln\left(\frac{T_{\rm GUT}}{T_{\rm HC}}\right)^4-1} \right)(\epsilon+\beta-2\eta),
\end{equation}
where $T_{\rm HC}$ is the temperature at the horizon crossing where the perturbation effectively occurs.
The spectral index \eqref{eq:ns2} is identical with the spectral index in Ref. \cite{Hall:2003zp}
except the last term  originated from the modified Stefan-Boltzmann law \eqref{eq:radrhop}.

\begin{figure}[hpb]
  \begin{center}
  \includegraphics[width=0.5\linewidth]{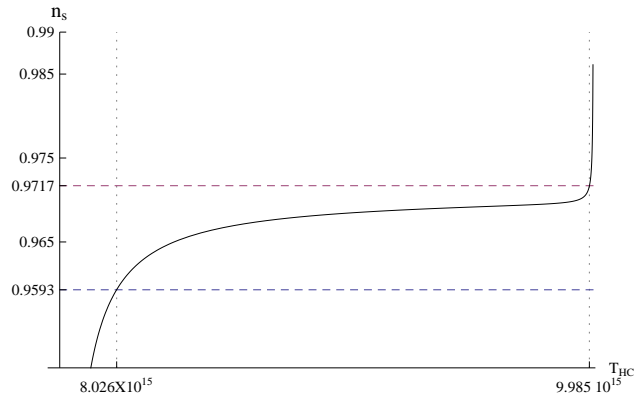}
  \end{center}
  \caption{The spectral index $n_{\rm s}$ vs the temperature at the horizon crossing $T_{\rm HC}$ is plotted
   such that the solid curve is the spectral index \eqref{eq:ns3}, where the number of e-folds and the GUT scale are fixed as $N_{\rm inf} = 60$ and $T_{\rm GUT} = 10^{16}{\rm GeV}$, and the two dashed lines show
    the range of the Planck 2015 data, $0.9593 \le n_{\rm s} \le 0.9717$.}
  \label{fig:nsTHC}
\end{figure}

Next, the unknown parameters such as $\lambda,~\phi_{\rm HC},~\Gamma_0,~\phi_0$
are eliminated by combining Eqs. \eqref{eq:DiffT}, \eqref{eq:NSR} with Eq. \eqref{eq:ns2}.
After some calculations,
the spectral index is finally obtained as
\begin{equation}\label{eq:ns3}
n_{\rm s}-1=
\frac{1}{ N_{\rm inf}+N_{\rm inf}^2 \ln\left(\frac{T_{\rm GUT}}{T_{\rm HC}}\right)^4} - \frac{1}{12 N_{\rm inf}\left(1-\ln\left(\frac{T_{\rm GUT}}{T_{\rm HC}}\right)^4\right)} -\frac{7}{4 N_{\rm inf}},
\end{equation}
where
the number of e-folds is assumed to be $N_{\rm inf} = 60$.
As shown in Fig. \ref{fig:nsTHC}, the spectral index can respect the data of Planck 2015
when the temperature at the horizon crossing $T_{\rm HC}$ lies in the interval of
$8.026 \times 10^{15} {\rm GeV} \le T_{\rm HC} \le 9.985 \times 10^{15} {\rm GeV}$.

To evaluate the temperature
at the end of warm inflation,
we take the procedure presented in Ref. \cite{Mielczarek:2010ag},
which was already applied to the non-minimal kinetic
coupling model \cite{Goodarzi:2014fna}.
By using Eq. \eqref{eq:SReq3}, the total number of e-folds $N_{\rm tot}$ from the scale at the horizon crossing $a_{\rm HC}$ to the scale at the present time $a_0$ is written as
\begin{align}\label{eq:N1}
N_{\rm tot} =\ln \left(\frac{a_0}{a_{\rm HC}} \right) =\ln \left(\frac{\sqrt{\lambda \phi_{\rm HC}^2-\gamma_{\rm HC}T_{\rm HC}^4 }}{\sqrt{3}k_0 m_{\rm p}}\right),
\end{align}
where the scale of the present time is fixed as
$a_0 = 1$ and the scale at the horizon crossing is given as $a_{\rm HC}=k_0/H(t_{\rm HC})$.

Next, the number of e-folds \eqref{eq:N1} can be divided into three parts composed of
inflationary regime $N_{\rm inf}= \ln (a_{\rm end}/a_{\rm HC})$,
 radiation-dominated era $N_{\rm rad}= \ln (a_{\rm rec}/a_{\rm end})$, and the time after recombination until now $N_{\rm 0}= \ln (a_{\rm 0}/a_{\rm rec})$ as
\begin{align}
N_{\rm tot}=N_{\rm 0}+N_{\rm rad}+N_{\rm inf}
=\ln \left(\frac{a_0}{a_{\rm rec}} \right)+\ln \left(\frac{a_{\rm rec}}{a_{\rm end}} \right)+\ln \left(\frac{a_{\rm end}}{a_{\rm HC}} \right), \label{eq:N2}
\end{align}
where $a_{\rm rec}$ and $a_{\rm end}$ are the scales at the recombination era and the end point of inflation, respectively.
The relation of
$T_{\rm rec}=(1+z_{\rm rec})T_{\rm CMB}$, where $z_{\rm rec}$ is the redshift factor given as $1+z_{\rm rec} = a_0/a_{\rm rec}$,
indicates that the temperature diminishes
from the recombination era to present universe
due to the expansion of the universe. So,
the first term in Eq. \eqref{eq:N2} can be expressed as
\begin{equation}\label{eq:Nfirst}
N_0=\ln \left(\frac{a_0}{a_{\rm rec}}\right) = \ln \left(\frac{T_{\rm rec}}{T_{\rm CMB}}\right).
\end{equation}
For the radiation-dominated era in Eq. \eqref{eq:N2},
the adiabatic expansion of the universe is assumed as $dS=0$ \cite{Kolb1990},
so that $S= a_{\rm rec}^3 s_{\rm rec}=a_{\rm end}^3 s_{\rm end}$.
Then the number of e-folds can be rewritten in the radiation-dominated era $N_{\rm rad}$
as
\begin{equation}\label{eq:Nsecond}
N_{\rm rad} =\ln \left(\frac{a_{\rm rec}}{a_{\rm end}} \right)=\frac{1}{3}\ln \left(\frac{s_{\rm end}}{s_{\rm rec}}\right)
=\frac{1}{3} \ln \left(\frac{4\gamma_{\rm end} T_{\rm end}^3 \ln\left(\frac{T_{\rm GUT}}{T_{\rm end}}\right)^4}{4\gamma_{\rm rec}T_{\rm rec}^3}  \right),
\end{equation}
where the entropy density at the end of inflation is
$s_{\rm end}=4\gamma_{\rm end} T_{\rm end}^3 \ln\left(T_{\rm GUT} /T_{\rm end}\right)^4$
from Eqs. \eqref{eq:spT} and \eqref{eq:ptot3}.
By the way, $s_{\rm rec}=4\gamma_{\rm rec}T_{\rm rec}^3$ since the radiation only consists of photons without
the inflaton, so that the usual Stefan-Boltzmann law is used.

Plugging Eqs. \eqref{eq:N1}, \eqref{eq:Nfirst}, \eqref{eq:Nsecond} into  Eq. \eqref{eq:N2},
we get
\begin{equation}\label{eq:phiHE2}
\ln \left(\frac{\sqrt{\lambda \phi_{\rm HC}^2-\gamma_{\rm HC}T_{\rm HC}^4 }}{\sqrt{3}k_0 m_{\rm p}}\right) = N_{\rm inf}+
\frac{1}{3} \ln \left(\frac{\gamma_{\rm end} T_{\rm end}^3 \ln\left(\frac{T_{\rm GUT}}{T_{\rm end}}\right)^4}{\gamma_{\rm rec}T_{\rm CMB}^3}  \right).
\end{equation}

To determine $T_{\rm end}$,
we choose the effective particle number at the electroweak energy scale as $g_{\rm HC}=g_{\rm end}=106.75$
and at the recombination era as $g_{\rm  rec}=2$ \cite{Kolb1990}.
The temperature of CMB is known as $ T_{\rm CMB}=2.725 K$, and
the spectral index for $k_{0} = 0.05 {\rm Mpc}^{-1}$
is $n_{\rm s}=0.9655 \pm 0.0062$
from Planck 2015 \cite{Ade:2015lrj, Ade:2015xua}.
In the previous section, the temperature $T_{\rm HC}$ at the horizon crossing was already evaluated
as $8.026 \times 10^{15} {\rm GeV} \le T_{\rm HC} \le 9.985 \times 10^{15}$
by solving Eq. \eqref{eq:ns3}.
After all, from Eq. \eqref{eq:phiHE2},
the range of $T_{\rm end}$ is obtained as
\begin{equation}\label{eq:Tem1}
2.409\times 10^{13} ~{\rm GeV} \le T_{{\rm end}} \le 2.216 \times 10^{14} ~{\rm GeV},
\end{equation}
where this range lies below the well-known upper bound of the temperature of the universe to
avoid monopole proliferation
~\cite{Kolb1990}
and above the lower bounds in Refs.~\cite{Kawasaki:2000en, Hannestad:2004px, Martin:2010kz}.
In addition, the corresponding energy density for radiation is consequently
\begin{equation}
2.852 \times 10^{56} {\rm GeV}^4 \le \rho_{\rm end} \le 1.291 \times 10^{60} {\rm GeV}^4,
\end{equation}
which is a sufficient radiation energy density to accommodate
the GUT baryogenesis at the end of inflation \cite{Bellini:2001ka}.

As a matter of fact,
we have assumed the simplest setting described by a perfect fluid as a toy model;
however, the decay process causes
the deviation of equilibrium and perfectness of the radiation as well as
the inflaton field. So, there might be some deviations from this limit,
which leads to viscous dissipation and corresponding noise forces.
On general grounds, random sources and dissipative stresses are introduced via a shear stress tensor $\Pi_{\mu\nu}$
in the energy-momentum tensor, $ T_{\mu\nu}=(\rho+p)u_\mu u_\nu +  g_{\mu\nu} p + \Pi_{\mu\nu}$ \cite{Bastero-Gil:2014jsa}.
According to Landau's theory of random fluids \cite{Landau1975},
the dissipation is governed by constitutive relations for shear viscosity $\eta_{\rm s}$ and bulk viscosity $\eta_{\rm b}$
 while fluctuations are generated by Gaussian noise term $\Sigma_{\mu\nu}$.
In a comoving frame, the non-vanishing shear terms are written as
$\Pi_{\mu\nu}=-(\eta_{\rm s} \nabla_\mu u_\nu + \eta_{\rm s} \nabla_\nu u_\mu +(\eta_{\rm b}-2\eta_{\rm s}/3)\delta_{\mu\nu} \nabla_\kappa u^\kappa)-\Sigma_{\mu\nu}$.
In this case, the energy-momentum tensor
is obtained as $T^\mu _\mu =-\rho +3p + \Pi^\mu_\mu$.
Since the shear terms $\Pi_{\mu\nu}$ are the traceless part of the energy-momentum tensor,
the trace of the shear terms $\Pi^\mu_\mu$ automatically vanishes for the radiation and inflaton field, respectively.
So, the total
trace for the radiation and inflaton is simply coincident with our trace relation in this work,
and then the shear terms $\Pi_{\mu\nu}$ consequently
do not affect the form of the modified Stefan-Boltzmann laws \eqref{eq:SBrho} and \eqref{eq:SBp} thanks to the traceless property of shear terms.
On the other hand, the effects due to the shear terms $\Pi_{\mu\nu}$ play a role in the cosmological perturbation
 as seen from Ref. \cite{Bastero-Gil:2014jsa}.
To investigate specific changes due to the shear terms, we need to calculate the cosmological perturbation for the imperfect fluid with the modified Stefan-Boltzmann laws  \eqref{eq:SBrho} and \eqref{eq:SBp},
which seems                                                                                                                                                                                                              to be a non-trivial task.

\section{Summary}
\label{sec:discussion8}
It was shown that the proper energy density which amounts to the curvature scale of
$ \sim 1/M^2$ could be obtained.
The explicit calculation for the two-dimensional soluble Schwarzschild black hole
showed that the radiation energy density exists in the free-fall frames and
it depends on both the free-fall positions and the black hole states.
In the Boulware state, the proper energy density is always negative divergent at the
horizon, which is independent of the initial free-fall positions.
For the Hartle-Hawking-Israel state, the proper energy density at the horizon is negative finite
when the free-fall toward the black hole begins at $r_s < r_c$ while it is
positive finite at the horizon for $r_s > r_c$, where $r_c=3M$ is the
critical free-fall position to determine the sign of the proper energy density
at the horizon. The Unruh state yields a slightly larger critical
value of $r_c \approx 3.1 M$ as compared to that of the Hartle-Hawking-Israel
state. The most important ingredient is that the proper energy density
in the Unruh state is divergent at the horizon \cite{Visser:1996ix,Eune:2014eka,Gim:2014swa}, which is
contradictory to the conventional results in Refs. \cite{Candelas:1980zt,Birrell:1982ix}
where the Kruskal coordinates are treated as the free-fall coordinates.
We resolved this conflict
by showing that
the Kruskal coordinates could not be local inertial coordinates on the future horizon
through the non-vanishing affine connections.

The common belief is that the Tolman temperature
is divergent at the horizon due to the infinite blueshift of the Hawking radiation.
However, the usual Stefan-Boltzmann law assuming the traceless stress tensor should be
consistently modified in order to accommodate the case where the stress
tensor is no longer traceless in the process of the Hawking radiation.
From the modified Stefan-Boltzmann
law, we obtained the effective Tolman temperature in the two and four dimensional
Schwarzschild black-hole backgrounds, and find that it is
finite everywhere outside the black hole horizon and vanishes at the horizon.
In fact, the vanishing effective Tolman temperature on the horizon
can be understood in terms of the Unruh effect~\cite{Unruh:1976db}, where the static
metric~\eqref{4Dmetric} near the horizon can be written by the Rindler
metric for a large black hole whose curvature scale is negligible.
The Unruh temperature is divergent in virtue of the infinite acceleration of the frame
where the fixed detector is very close to the horizon.
So the Unruh temperature is equivalent to the
fiducial temperature for the Schwarzschild black hole~\cite{Singleton:2011vh},
which means that the geodesic observer should find the vanishing temperature on the
horizon since the proper acceleration of the geodesic detector vanishes.
In addition to this, AMPS argument \cite{Almheiri:2012rt} is that the firewall
on the horizon should be defined in an
evaporating black hole rather than the black hole in thermal equilibrium.
Using the advantage of the effective Tolman temperature, we
find the reason why the firewall could not exist in thermal equilibrium.

The conventional Tolman temperature in the Unruh vacuum
leads to the fact that
Hawking radiation at infinity comes
from the infinitely blueshifted outgoing Hawking excitations at the horizon;
however, it was a misleading interpretation due to the two overlapped features.
This issue was clarified by decomposing the Tolman temperature in the Unruh vacuum
into the left and right chiral temperatures.
We showed that the firewall
in the Unruh vacuum comes from the negative influx crossing the horizon,
while the Hawking radiation in the Unruh vacuum
comes from the positive
outward flux in the near-horizon quantum region of the atmosphere, not right at the horizon.
The right temperature is finite everywhere, which means that
the low energy Hawking particles are irrelevant to the infinite blueshift.
The firewall from the infinite Tolman temperature at the horizon
and the Hawking radiation from the atmosphere outside the horizon are compatible,
when we discard the fact that the Hawking radiation in the Unruh vacuum
originates from the infinitely blueshifted
outgoing excitations at the horizon.
In connection with the information loss paradox, the firewall need not play
a role of the entanglement-breaker between two partners of each pair across the horizon
created from pair production \cite{Israel:2015ava}. The origin of the firewall is just
due to the infinitely blueshifted influx crossing the horizon \cite{Kim:2016iyf}.
If the firewall could be regularized in certain ways,
then it would be possible to save the violation of the equivalence principle at the horizon.
Of course, there may be other kinds of resolutions, such that
there are no event horizons \cite{Hawking:2014tga} and
the event horizons are inappropriate to describe the evaporating
black hole \cite{Visser:2014zqa}; otherwise,
the quantum back reaction of the geometry
renders a star stop collapsing a finite radius larger than
its horizon \cite{Mersini-Houghton:2014zka, Mersini-Houghton:2014cta}.
Recently, it was also claimed that the firewall is due to the limitation of
the semiclassically fixed background
in that the semiclassical theory possesses an unphysically large Fock space
built by creation and annihilation operators on a fixed black hole background \cite{Nomura:2016qum}.
This issue deserves further attention.

Finally,
motivated by the non-zero initial radiation energy density in warm inflation scenario,
we performed thermodynamic analysis for the warm inflation model
by using the definitions for the inflaton and radiation energy density presented in Ref. \cite{Kolb1990}
and obtained the modified Stefan-Boltzmann law
to show that the zero radiation energy density \eqref{eq:radrhop} at
the Grand Unification epoch just prior to starting inflation became finite when inflation starts,
which gives the adequate radiation energy density for warm inflation.
By using the effective Tolman temperature from the modified Stefan-Boltzmann law for the radiation energy density,
we studied the number of e-folds and the spectral index of the scalar perturbation under the slow-roll approximations
in the power-law potential and damping terms,
so that
the temperature \eqref{eq:Tem1} at the end of warm inflation was successfully calculated,
and it satisfies the upper bound lower than the GUT scale \cite{Kolb1990},
and lower bound of the big bang nucleosynthesis~\cite{Kawasaki:2000en, Hannestad:2004px}
by the CMB data \cite{Martin:2010kz}.
Additionally,  we confirmed that
a sufficient radiation energy density could be produced for GUT baryogenesis at the end of inflation \cite{Bellini:2001ka}.

In conclusion, it has been shown that
the effective Toman temperature derived
from the modified Stefan-Boltzmann law could resolve many interesting gravitational and cosmological problems.

\acknowledgments

I have been benefited from discussion with
Myungseok Eune and Youngwan Gim.
This work was supported by the
National Research Foundation of Korea (NRF) grant funded by the
Korea government (MSIP) (2017R1A2B2006159).
\bibliographystyle{JHEP}       

\bibliography{references}


\end{document}